\documentclass[a4paper,prd,showpacs,noshowkeys]{revtex4}
\usepackage[latin1]{inputenc}
\usepackage{amsmath}
\usepackage{amssymb}
\usepackage{mathrsfs}

\providecommand{\bx}{{\bf x}}
\providecommand{\bv}{\mathbf{v}}
\providecommand{\bbA}{\mathbb{A}}
\providecommand{\bbB}{\mathbb{B}}
\providecommand{\tp}{\!\mt{\top}}

\providecommand{\ca}{\mathcal{A}}
\providecommand{\cs}{\mathcal{S}}
\providecommand{\ft}{\mathfrak{t}}
\providecommand{\cG}{\mathcal{G}}
\providecommand{\ad}{\mathrm{ad}}
\providecommand{\diag}{\mathrm{diag}}
\def\Tr{\mathrm{Tr}}
\def\bM{\mathbf{M}}

\providecommand{\mn}[1]{\mbox{\normalsize $#1$}}
\providecommand{\ms}[1]{\mbox{\small $#1$}}
\providecommand{\mt}[1]{\mbox{\tiny $#1$}}
\providecommand{\ml}[1]{\mbox{\large $#1$}}
\providecommand{\undertilde}[1]{\underset{\mn{\tilde{}}}{#1}}
\providecommand{\eq}[1]{\begin{equation} #1 \end{equation}}
\providecommand{\eqarr}[1]{\begin{eqnarray} #1 \end{eqnarray}}
\providecommand{\bs}[1]{\boldsymbol{#1}}
\providecommand{\ket}[1]{\vert #1 \rangle}
\providecommand{\bra}[1]{\langle #1 \vert}
\providecommand{\aver}[1]{\langle #1 \rangle}
\providecommand{\hs}[1]{\hspace{#1}}

\font\bb=bbmss12 scaled 1000
\def\id{\mbox{\bb 1}}

\begin{document}
\title{CP violation conditions in N-Higgs-doublet potentials}
\author{C.~C.~Nishi}
\email{ccnishi@ift.unesp.br}
\affiliation{Instituto de F\'\i sica Te\'orica,
UNESP -- São Paulo State University\\
Rua Pamplona, 145\\
01405-900 -- S\~ao Paulo, Brazil
}

\begin{abstract}
Conditions for CP violation in the scalar potential sector of general
N-Higgs-doublet models (NHDMs) are analyzed from a group theoretical
perspective. For the simplest two-Higgs-doublet model (2HDM) potential, a
minimum set of conditions for explicit and spontaneous CP violation is
presented. The conditions can be given a clear geometrical interpretation in
terms of quantities in the adjoint representation of the basis transformation
group for the two doublets. Such conditions depend on CP-odd pseudoscalar
invariants. When the potential is CP invariant, the explicit procedure to reach
the real CP-basis and the explicit CP transformation can also be obtained. The
procedure to find the real basis and the conditions for CP violation are then
extended to general NHDM potentials. The analysis becomes more involved and only
a formal procedure to reach the real basis is found. Necessary conditions for CP
invariance can still be formulated in terms of group invariants: the CP-odd
generalized pseudoscalars. The problem can be completely solved for three
Higgs-doublets.

\end{abstract}
\maketitle
\section{Introduction}
\label{sec:intro}

It is well known that group automorphisms play an important role in the
CP violation phenomenon. In a extensive paper, Grimus and Rebelo\;\cite{GR} have
analyzed the CP-type transformations as automorphisms in the gauge symmetry
present in the Quantum Field Theory (QFT) models of particle physics. They
showed, at the classical level, that general gauge theories with fermions and
scalars coupled to gauge bosons through minimal coupling are always CP
invariant. In other words, a CP-type transformation that is a symmetry of the
theory can always be found. The only terms that can possibly violate the CP
symmetry are the Yukawa couplings and scalar potentials. In the Standard Model
(SM), the unique source of CP violation comes from the complex phases in the
Yukawa couplings that are transferred to the Cabibbo-Kobayashi-Maskawa (CKM)
matrix\;\cite{KM} after spontaneous electroweak symmetry breaking (EWSB). Within
such context the possibility of (explicit) CP violation is intimately connected
with the presence of a horizontal space: the quarks come in three identical
families distinguished only by their masses.

Another source of CP violation could arise in the scalar potential
sector\;\cite{branco:book}. In such case two patterns can be possible, either
the CP symmetry is violated explicitly in the theory before EWSB or the CP
violation arises spontaneously jointly with EWSB. Several models with
 spontaneous
CP violation arising from the Higgs sector were constructed after
Refs.\;\onlinecite{Lee} and \onlinecite{Weinberg}, aiming to attribute to the
violation of CP the same origin of the broken-hidden gauge symmetries.
Nonetheless, the available CP violation data seems to be in general accordance
with the SM CKM mechanism\;\cite{PDG,ckmfitter}. Then, concerning the CP
violation data, restricted by mixing constraints and strong suppresion of flavor
changing neutral currents (FCNC)\;\cite{GW}, the challenge is to develop a model
that incorporates entirely or partially the CKM mechanism.

The scalar potential sector, although phenomenologically rich in
CP violating sources (see, {\it e.g.}, Refs.\;\onlinecite{Weinberg} and
\onlinecite{WW}), has not yet been analyzed for general gauge theories under a
group theoretical perspective. One of the reasons for the difficulty for a
general treatment is that the scalar potential involves higher order
combinations of scalar fields than other sectors of gauge theories, with terms
constrained only by the underlying gauge symmetries and, if required,
renormalizability. Renormalizability in four dimensions constrains the highest
order scalar field combination to be quartic.

Another difficulty for analyzing the CP violation properties for general
gauge theories is the freedom to change the basis of fields used to describe the
theory. The most familiar is the SM's rephasing freedom for the quark fields:
this change of basis transforms the CKM matrix and the complex entries.
Such ambiguity can be avoided by using rephasing invariants\;\cite{jarlskog}
which depend only on one physically measurable CP phase.
More generally, for theories with horizontal spaces, there is a freedom to
continuously rotate the basis of such spaces without changing the physical
content. For this case, it is also possible to write the observables in a
basis independent
manner\;\cite{botella:94,davidson,Haber.2,GH,ivanov:05,branco:05}, or, in other
words, in terms of reparameterization invariants\;\cite{ginzburg}. In any case,
it is important to be able to establish general conditions for CP violation to
analyze more transparently the possible CP violating patterns for gauge theories
with large gauge groups and/or horizontal spaces.

Following this spirit of classifying and quantifying CP violation based on basis
invariants, it will be treated here the simplest class of extensions of the SM:
the multi-Higgs-doublet models\;\cite{branco:79,mendez:91,lavoura:94,nhdm},
which we shall denote by NHDM for $N$ Higgs-doublets. The simplest of them is
the two-Higgs-doublet model (2HDM) which has been extensively
studied in the literature\;\cite{WW,carena,wu}, also employing the basis
independent methods\;\cite{GH,davidson,Haber.2,branco:05,ivanov:05}. An explicit
but not complete study for 3HDM potentials can be found in
Ref.\;\onlinecite{branco:05}. The recent interest is based on the fact that the
2HDM can be considered as an effective theory of the minimal supersymmetric
extension of the SM (MSSM)\;\cite{carena}, which requires two Higgs-doublets
from supersymmetry.

Concerning the 2HDMs, a throughout analysis of the CP symmetry aspect of
the 2HDM potentials was presented recently\;\cite{GH,ivanov:05}. The
necessary and sufficient conditions for spontaneous and explicit CP
violation were presented, expressed in terms of basis independent conditions and
invariants. In this respect, in Sec.\;\ref{sec:N=2}, a more compact version of
such proofs will be shown. The approach used is much alike the
one presented in Ref.\;\onlinecite{ivanov:05}: from group theoretical analysis,
the adjoint representation can be used as the minimum nontrivial representation
of the transformation group of change of basis for the two doublets, i.e., the
horizontal $SU(2)$ group. Working with the adjoint representation allows for an
alternative formulation of the CP invariance conditions which facilitate the
analysis and enables one, when the potential is CP invariant, to find the
explicit CP transformation and the explicit {\it real basis}\;\cite{davidson},
i.e., the basis for which all the parameters in the potential are real. Such
issues were not addressed in previous approaches\;\cite{GH,branco:05,ivanov:05}.
The basis independent conditions are formulated in terms of pseudoscalars of the
adjoint. In Sec.\;\ref{subsec:scpv:N=2}, we also obtain the necessary and
sufficient conditions to have spontaneous CP violation.

In Sec.\;\ref{sec:N>2}, an extension of the method is attempted
to treat general NHDMs. The analysis becomes much more involved than the $N=2$
case and further mathematical machinery is necessary. Nevertheless, stringent
necessary conditions for CP invariance can be formulated. Generalized
pseudoscalars, which should be null for a CP invariant potential, can still be
constructed. For $N=3$, the conditions found are shown to be sufficient if
supplemented by an additional condition.
In Sec.\;\ref{subsec:scpv:N>2}, a brief account on spontaneous CP violation on
NHDMs is presented.

At last, in Sec.\;\ref{sec:concl} we draw some conclusions and discuss some
possible approaches for the complete classification of the CP-symmetry
properties for the NHDMs. (Some useful material is also presented in the
appendices.)

\section{$N=2$ Higgs-doublets}
\label{sec:N=2}

For $N=2$ Higgs-doublets $\Phi_a$, $a=1,2$, transforming under
$SU(2)_L\otimes U(1)_Y$ as $(2,1)$, the minimal gauge invariant
combinations that can be constructed are
\eq{
\label{inv:N=2}
A_a=\Phi_a^\dag \Phi_a~,~~a=1,2,
~~B=\Phi_1^\dag\Phi_2 \text{ and } B^\dag~.
}
All other invariants can be constructed as combinations of these
ones\,\cite{endnote1}. Thus the most general renormalizable 2HDM
potential can be parameterized as\;\cite{GH}
\eqarr{
\label{V:N=2}
V(\Phi)&=&
m^2_{11}A_1+m^2_{22}A_2-(m^2_{12}B+h.c.)+
\frac{\lambda_1}{2}A^2_1+\frac{\lambda_2}{2}A^2_2+
\lambda_3A_1A_2+\lambda_4BB^\dag
\cr
&&~+
\{
\frac{\lambda_5}{2}B^2+[\lambda_6A_1+\lambda_7A_2]B+h.c.\}
~.
}
From the hermiticity condition,
$\{m^2_{11}, m^2_{22}, \lambda_1, \lambda_2, \lambda_3, \lambda_4\}$ are real
parameters and $\{m^2_{12}, \lambda_5, \lambda_6, \lambda_7\}$ are potentially
complex, summing up to $6+2\times 4=14$ real parameters.

The existence of complex parameters {\it per se}, though, does not mean the
potential in Eq.\;\eqref{V:N=2} is CP violating. A $U(2)_H$ horizontal
transformation can be performed on the two doublets possessing identical quantum
numbers, except, perhaps, for their masses,
\eq{
\label{horizontal:N=2}
\Phi
\rightarrow
U \Phi
}
where $U$ is a $U(2)$ transformation matrix and
$\Phi$ denotes the assembly of the two doublets in
\eq{
\label{Phi:N=2}
\Phi\equiv
\begin{pmatrix}
\Phi_1\cr \Phi_2
\end{pmatrix}
~.
}
Actually, the global phase transformation in $U$ amounts for a hypercharge
transformation under which the gauge invariants in Eq.\;\eqref{inv:N=2} do
not change. Thus only a $U\in SU(2)_H$ transformation needs to be considered.
This basis transformation freedom suggests, and indeed it can be
proved\;\cite{branco:book,GH}, that the necessary and sufficient conditions for
$V(\Phi)$ to be CP invariant are equivalent to the existence of a basis
reached by a transformation $U$ \eqref{horizontal:N=2} in which all the
parameters present in the potential are real. Since these basis transformations
can be reformulated as transformations on the parameters, all the analysis
resumes in investigating the transformation properties of the parameters of
$V(\Phi)$ under $SU(2)_H$. Indeed, the parameters can be written as higher order
tensors, transforming under the fundamental representation of
$SU(2)_H$\;\cite{davidson,GH,Haber.2}.

Instead of performing the analysis of tensors under the fundamental
represention $\mathbf{2}$ of $SU(2)_H$, as in Ref.\,\onlinecite{GH}, we can
take advantage of the form of the minimal invariants \eqref{inv:N=2} that
transform as $\bar{\mathbf{2}}\otimes \mathbf{2}=\mathbf{3}\oplus \mathbf{1}$,
 by
using as the minimal nontrivial representation the adjoint $\mathbf{3}$. In fact
this property does not depend on the number of doublets $N$, and the
invariants of the type of Eq.\;\eqref{inv:N=2} always form representations of
$SU(N)_H$ with decomposition
\eq{
\label{branching1}
\bar{\mathbf{N}}\otimes \mathbf{N}=\text{adj}\oplus \mathbf{1}~.
}
This property will  be exploited in section \ref{sec:N>2} to treat general
$N$-Higgs-doublet potentials.

For $SU(2)_H$, the decomposition in Eq.\;\eqref{branching1} can be performed
by using instead of $\{A_1,A_2,B,B^\dag\}$ the real combinations
\eq{
\label{A:N=2}
\bbA_\mu\equiv
\mn{\frac{1}{2}}\Phi^\dag\sigma_\mu \Phi
~,~~\mu=0,1,\ldots,3,
}
where $\sigma_\mu=(\id,\bs{\sigma})$. The Greek index is not a space-time index,
which means there is no distinction between covariant or contravariant
indices but the convention of summation over repeated indices will be used.
The indices running over $\mu=i=1,2,3$ are group indices in the
space of the Lie algebra, i.e., in the adjoint representation and the $\mu=0$
index is the trivial singlet component.
The explicit change of basis reads
\eqarr{
\label{A->Amu:N=2}
\bbA_0&=&\frac{A_1+A_2}{2}~,\cr
\bbA_3&=&\frac{A_1-A_2}{2}~,\cr
\bbA_1&=&\frac{B+B^\dag}{2}=
\mathrm{Re}B~,\cr
\bbA_2&=&\frac{B-B^\dag}{2i}=
\mathrm{Im}B~,
}
which can be readily inverted and inserted in the potential of
Eq.\;\eqref{V:N=2} to give compactly
\eq{
\label{V2:N=2}
V(\bbA)=M_\mu\bbA_\mu+\bbA_\mu\Lambda_{\mu\nu}\bbA_\nu~,\\
}
where
\eqarr{
\{M_\mu\}&=&
(m^2_{11}+m^2_{22},-2\mathrm{Re}\,m^2_{12},2\mathrm{Im}\,m^2_{12},
m^2_{11}-m^2_{22}
)
\\
\label{Lambda:N=2}
\Lambda&=&\{\Lambda_{\mu\nu}\}=
\left(
\begin{array}{c|ccc}
\bar{\lambda}+\lambda_3 & \mathrm{Re}(\lambda_6+\lambda_7) &
-\mathrm{Im}(\lambda_6+\lambda_7) & \Delta \lambda/2
\cr\hline
\mathrm{Re}(\lambda_6+\lambda_7)& \lambda_4+\mathrm{Re}\lambda_5&
-\mathrm{Im}\lambda_5 &\mathrm{Re}(\lambda_6-\lambda_7)
\cr
-\mathrm{Im}(\lambda_6+\lambda_7) & -\mathrm{Im}\lambda_5&
\lambda_4-\mathrm{Re}\lambda_5&-\mathrm{Im}(\lambda_6-\lambda_7)
\cr
\Delta \lambda/2 & \mathrm{Re}(\lambda_6-\lambda_7)&
-\mathrm{Im}(\lambda_6-\lambda_7)&\bar{\lambda}-\lambda_3
\end{array}
\right)
~,
}
and $\bar{\lambda}=(\lambda_1+\lambda_2)/2$,
$\Delta\lambda=\lambda_1-\lambda_2$. Notice that all parameters in this basis
are real and the criterion for CP violation have to be different of the reality
condition. Furthermore, $\Lambda$ is real and symmetric.

The coefficients of $M$ can be more conveniently written as
\eq{
\label{M:N=2}
M_\mu \equiv \Tr[\sigma_\mu Y]~,
}
where
\eq{
Y=
\begin{pmatrix}
m^2_{11} & - m^2_{12}\cr
- m^{2*}_{12}& m^2_{22}
\end{pmatrix}
=M_\mu \mn{\frac{1}{2}}\sigma_\mu
~,
}
is the mass matrix for
\eq{
V(\Phi)\Big|_{\Phi^2}=\Phi^\dag Y\Phi~.
}
These relations can be easily extended to general $N$ doublets by replacing
the $\{\sigma_i\}$ matrices by the proper generators of $SU(N)_H$,
$\{\lambda_i\}$, and the corresponding identity matrix. The relation
\eqref{M:N=2} follows from the completeness of the basis
$\{\sigma_\mu\}$ in the space of complex $2\times 2$ matrices\;\cite{fierz}.

Expanding Eq.\;\eqref{V2:N=2} in terms of the irreducible pieces of
$\mathbf{3}\oplus\mathbf{1}$,
\eq{
\label{V3:N=2}
V(\bbA)=M_0\bbA_0+\Lambda_{00}(\bbA_0) ^2+M_i\bbA_i
+2\Lambda_{0i}\bbA_0\bbA_i
+\bbA_i\tilde{\Lambda}_{ij}\bbA_j~,
}
we identify two vectors
$\mathbf{M}\equiv \{M_i\},\bs{\Lambda}_0\equiv \{\Lambda_{0i}\}$ and one rank
two tensor $\tilde{\Lambda}=\{\Lambda_{ij}\}$ with respect to $\mathbf{3}$.
Further mention to the representation will be suppressed and it will be assumed
that the representation in question is the adjoint if otherwise not stated.
(For example, ``vectors'' and ``tensors''  transform under the adjoint.)
The tensor $\tilde{\Lambda}$ can be further reduced into irreducible pieces
following $\mathbf{3}\otimes\mathbf{3}=\mathbf{5}\oplus \mathbf{1}$ as
\eq{
\tilde{\Lambda}=\tilde{\Lambda}^{(0)}+\tilde{\Lambda}^{(5)}~.
}
The singlet component is just
$\tilde{\Lambda}^{(0)}=\mn{\frac{1}{3}}\Tr[\tilde{\Lambda}]\,\id_3$,
and the remaining of $\tilde{\Lambda}$ is the $\mathbf{5}$-component.
This last decomposition of $\tilde{\Lambda}$, though, will not be necessary
for the analysis because of the particular fact that the adjoint of $SU(2)\sim
SO(3)$ and all analysis can be done considering the rotation group in three
dimensions, which is very much known. The $SU(2)\rightarrow SO(3)$ two-to-one
mapping is given by the transformation induced by Eq.\;\eqref{horizontal:N=2}
over the invariants $\bbA_\mu$, \eqarr{
\bbA_0&\rightarrow& \bbA_0\cr
\bbA_i &\rightarrow& O_{ij}(U)\bbA_{j}
~,
}
where
\eq{
O_{ij}(U)\equiv \mn{\frac{1}{2}}\Tr[U^\dag\sigma_i U\sigma_j]
~, ~~\in SO(3)~.
}
If $U=\exp(i\bs{\sigma}\cdot\bs{\theta}/2)$,
$O_{ij}(\bs{\theta})=[\exp(i\theta_kJ_k)]_{ij}$,
where $(iJ_k)_{ij}=\varepsilon_{kij}$ are the generators of $SU(2)$
[$SO(3)$] in the adjoint representation.

At this point we have to introduce the transformation properties of the
scalar doublets under CP. One possible choice is
\eq{
\label{CP:Phi:N=2}
\Phi_a(x)\stackrel{CP}{\longrightarrow}
\Phi_a^*(\hat{x})~,
}
where $\hat{x}=(x_0,-\bx)$ for $x=(x_0,\bx)$.
The transformation of Eq.\;\eqref{CP:Phi:N=2} induces in the invariants
$\bbA_\mu$ the transformation
\eqarr{ \label{CP:A:N=2}
\bbA_0(x)&\stackrel{CP}{\longrightarrow}&\bbA_0(\hat{x}) \cr
\bbA_i(x)&\stackrel{CP}{\longrightarrow}&(I_2)_{ij}\bbA_j(\hat{x})
~,
}
where $I_2\equiv \mathrm{diag}(1,-1,1)$ represents the reflection in the 2-axis.
We shall denote the transformations \eqref{CP:Phi:N=2} and \eqref{CP:A:N=2} as
canonical CP-transformations and, in particular, the second equation of
Eq.\;\eqref{CP:A:N=2} as the canonical CP-reflection.
Since horizontal transformations are also allowed, the most general CP
transformation is given by the composition of Eqs.\;\eqref{CP:Phi:N=2} and
\eqref{CP:A:N=2} with $SU(2)_H$ transformations; any additional phase can be
absorbed in those transformations.
Thus the CP transformation over $\bbA_i$ involves a
reflection and it does not belong to the proper rotations $SO(3)$ induced by
horizontal transformations.
The question of CP invariance, then, resumes in the existence of horizontal
transformations composite with a reflection that leaves the potential invariant.
Since the reflection  along the 2-axis can be transformed into the reflection
along any axis through the composition with rotations, the natural choice of
basis is the basis for which $\tilde{\Lambda}$ is diagonal:
$ O^{CP}\tilde{\Lambda} O^{CP^{\tp}}=\diag(\{\tilde{\lambda}_i\})$. It is always
possible to find $O^{CP} \in SO(3)$ because $\tilde{\Lambda}$ is real and
symmetric. Furthermore, $O^{CP}$ is unique up to reordering of the diagonal
values $\{\tilde{\lambda}_i\}$, or up to rotations in the subspace of
 degenerate
eigenvalues in case $\{\tilde{\lambda}_i\}$ are not all different. In such
basis, $\bbA_i \rightarrow \bbA'_i=O^{CP}_{ij}\bbA_j$, the potential in
Eq.\;\eqref{V3:N=2} becomes
\eq{
\label{V4:N=2}
V(\bbA)=M_0\bbA_0+\Lambda_{00}(\bbA_0)^2+M'_i\bbA'_i
+2\Lambda'_{0i}\bbA_0\bbA'_i
+\tilde{\lambda}_i(\bbA'_i)^2~,
}
where
\eq{
\mathbf{M}'=O^{CP}\mathbf{M}
~\text{ and }~
\bs{\Lambda}_0'=O^{CP}\bs{\Lambda}_0~.
}
The last term of Eq.\;\eqref{V4:N=2} is reflection invariant along any of the
principal axes of $\tilde{\Lambda}$, $e'_1, e'_2, e'_3$
(if $\tilde{\Lambda}$ does not have degenerate eigenvalues,
the three principal axes are the only directions leaving the tensor
invariant by reflection; with degeneracies, a continuous set of directions exist
in the degenerate subspace). The only terms that must be considered are the
third and the fourth ones. They depend on two vectors $\mathbf{M}'$ and
$\mathbf{\Lambda}'_0$: for the potential in Eq.\;\eqref{V4:N=2} to be invariant
by reflection, and consequently by CP, it is necessary that the vectors
$\mathbf{M}'$ and $\mathbf{\Lambda}'_0$ be null for the same  component. In such
case, it is always possible by a suitable $\pi/2$ rotation to choose that
direction to be the 2-axis. This is the canonical CP-basis (the {\it real basis}
in Ref.\;\onlinecite{GH}) which have all the parameters in the potential
$V(\Phi)$ real, since there is no $\bbA'_2$ components, which are the only
possible source of complex entries in the change of basis of
Eq.\;\eqref{A->Amu:N=2}. The CP transformation in terms of the original
variables is recovered with the inverse transformation as
\eqarr{
\label{CP:U:N=2} \Phi_a(x)&\stackrel{CP}{\longrightarrow}&
(U^{CP^{\dag}}U^{CP^{*}})_{ab}\Phi_b^*(\hat{x})~,\cr
\bbA_i(x)&\stackrel{CP}{\longrightarrow}&(O^{CP^{\tp}}I_2O^{CP})_{ij}
\bbA_j(\hat{x}) ~, } where $O^{CP}=O(U^{CP})$.

The conditions we have found rely on a systematic procedure to find the
canonical CP-basis. The basis may not exist and the theory is CP violating.
In this two doublet case, the change of basis can be easily achieved by a
diagonalization. However, sometimes it is more useful to have a direct criterion
to check if the CP invariance holds before going to the procedure of finding
the CP-basis. For $N>2$ doublets the procedure of finding the CP-basis is not
straightforward and direct criteria are much more helpful.
The criteria for $N=2$ can be formulated with the pseudoscalar invariants
\eq{
\label{pseudoS}
I({\bv_1,\bv_2,\bv_3})=\varepsilon_{ijk}v_{1i}v_{2j}v_{3k}=
(\mathbf{v}_1\times \mathbf{v}_2)\cdot\mathbf{v}_3~.
}
It is common knowledge that the pseudoscalars defined by Eq.\;\eqref{pseudoS}
are invariant by rotations but changes sign under a reflection or a
space-inversion. Consequently, if the potential $V(\bbA)$ is reflection
invariant (CP-invariant), then all pseudoscalar invariants of the theory are
null. The lowest order non-trivial pseudoscalars that can be constructed with
two vectors $\{\mathbf{M},\bs{\Lambda}_0\}$ and one rank-2 tensor
$\tilde{\Lambda}$ are
\eqarr{
\label{Is:N=2}
I_M&=&I(\mathbf{M},\tilde{\Lambda}\mathbf{M},\tilde{\Lambda}^2\mathbf{M})~, ~~
\\
\label{IL0}
I_{\Lambda_0}&=&
I(\bs{\Lambda}_0,\tilde{\Lambda}\bs{\Lambda}_0,\tilde{\Lambda}^2\bs{\Lambda}_ 0)
 ~, ~~
\\
I_1&=&
I(\mathbf{M},\bs{\Lambda}_0,\tilde{\Lambda}\mathbf{M})~, ~~
\\
\label{I2}
I_2&=&
I(\mathbf{M},\bs{\Lambda}_0,\tilde{\Lambda}\bs{\Lambda}_0)~,
}
with dimensions $M^3\Lambda^3, \Lambda^6, M^2\Lambda^2$ and $M\Lambda^3$
respectively.

The following statements will be proved:
\begin{itemize}
\item[(A)] If $\mathbf{M}\times\bs{\Lambda}_0\neq 0$, $I_1=0$ and $I_2=0$ are
the necessary and sufficient conditions to $V(\bbA)$  be CP invariant. The CP
reflection direction is $\mathbf{M}\times\bs{\Lambda}_0$ and it is also an
eigenvector of $\tilde{\Lambda}$.

\item[(B)] If $\mathbf{M}\parallel \bs{\Lambda}_0$, $I_M=0$ (or
$I_{\Lambda_0}=0$) is the necessary and sufficient condition to $V(\bbA)$ be
CP invariant. The CP reflection direction is either $\mathbf{M}\times
\tilde{\Lambda}\mathbf{M}$ ($\neq 0$) or an eigenvector of $\tilde{\Lambda}$
perpendicular to $\mathbf{M}$ (if $\tilde{\Lambda}\mathbf{M}\parallel
\mathbf{M}$) and the CP reflection direction is an eigenvector of
$\tilde{\Lambda}$.

\item[(C)] All higher order pseudoscalar invariants are null if (A) or (B) is
true.
\end{itemize}

The statements (A) and (B) are proved by noting that $I(\bv_1,\bv_2,\bv_3)=0$
implies that $\bv_1, \bv_2, \bv_3$ lie in the same plane.
For (A), if $I_1=I_2=0$, we can write $\tilde{\Lambda}\mathbf{M}=\alpha
\mathbf{M}+\beta\bs{\Lambda}_0$ and $\tilde{\Lambda}\bs{\Lambda}_0=\alpha'
\mathbf{M}+\beta'\bs{\Lambda}_0$, which means the application of
$\tilde{\Lambda}$ on $\mathbf{M}$ or $\bs{\Lambda}_0$ lie on the plane
perperpendicular to $\mathbf{M}\times\bs{\Lambda}_0$;
$I_M=0$ and $I_{\Lambda_0}=0$ are automatic.
Then $I_M=0$ implies $\tilde{\Lambda}^2\mathbf{M}=\alpha''
\mathbf{M}+\beta''\tilde{\Lambda}\mathbf{M}$, which means
$\tilde{\Lambda}^n\mathbf{M}$ remains in the plane defined by
$\{\mathbf{M},\bs{\Lambda}_0\}$. The same reasoning apply to
$\tilde{\Lambda}^m\bs{\Lambda}_0$ from $I_{\Lambda_0}=0$.
Then, the set $\{\mathbf{M},\bs{\Lambda}_0\}$ defines a principal plane of
$\tilde{\Lambda}$, i.e., a plane perpendicular to a principal axis
of $\tilde{\Lambda}$, the vector $\mathbf{M}\times\bs{\Lambda}_0$, which is then
an eigenvector of $\tilde{\Lambda}$. The latter can be seen from
$I_1=(\mathbf{M}\times\bs{\Lambda}_0).(\tilde{\Lambda}\mathbf{M})
=(\tilde{\Lambda}(\mathbf{M}\times\bs{\Lambda}_0)).\mathbf{M}=0$, and
$\tilde{\Lambda}(\mathbf{M}\times\bs{\Lambda}_0)$ is perpendicular to
$\mathbf{M}$; analogously $I_2=0$ implies
$\tilde{\Lambda}(\mathbf{M}\times\bs{\Lambda}_0)$ is also perpendicular to
$\bs{\Lambda}_0$, therefore
$\tilde{\Lambda}(\mathbf{M}\times\bs{\Lambda}_0)\propto
(\mathbf{M}\times\bs{\Lambda}_0)$.
At last, choose $(\mathbf{M}\times\bs{\Lambda}_0)$ as the reflection direction
($e'_2$-axis), then $\mathbf{M}$ and $\bs{\Lambda}_0$ have null projection
with respect to that direction and the CP-basis is found. This proves that
$I_1=I_2=0$ is a sufficient condition. That it is also necessary, can be seen
through the search of the CP-basis: a CP-basis requires both
$\{\mathbf{M},\bs{\Lambda}_0\}$ to be in  the same principal plane, then
$\tilde{\Lambda}^n\mathbf{M}$ or $\tilde{\Lambda}^m\bs{\Lambda}_0$ remain
in that plane and $I_1=I_2=0$.

For the disjoint case (B), $I_1=I_2=0$ is automatic. There is only one
independent direction and a rank-2 tensor. $I_M=0$ (or $I_{\Lambda_0}=0$)
implies that either $\tilde{\Lambda}\mathbf{M}\parallel \mathbf{M}$ and
$\mathbf{M}$ is an eigenvector, or $\tilde{\Lambda}^n\mathbf{M}=
\alpha\mathbf{M} + \beta\tilde{\Lambda}\mathbf{M}$ and
$\{\tilde{\Lambda}\mathbf{M},\mathbf{M}\}$ defines a principal plane.
Then, use either $\mathbf{M}\times\tilde{\Lambda}\mathbf{M}$ ($\neq 0$)  or an
eigenvector of $\tilde{\Lambda}$ perpendicular to $\mathbf{M}$
($\tilde{\Lambda}\mathbf{M}\parallel \mathbf{M}$) as the
CP-reflection direction and the CP-basis is achieved. The converse is also true,
if a CP-basis can be found, the invariants are null.

A subtlety arises when $\tilde{\Lambda}$ have degeneracies. When only two
eigenvalues are equal, still one principal direction and a perpendicular
principal plane is defined; every vector in the latter plane is an
eigenvector and any plane containing the non-degenerate eigenvector is also a
principal plane. With these extended definitions the proofs above are still
valid. For the trivial case when the three eigenvalues are degenerate,
$\tilde{\Lambda}$ is proportional to the identity and a CP-basis can always be
found by using $\mathbf{M}\times\bs{\Lambda}_0$ as the CP-reflection direction.
It is also important to remark that only for $\tilde{\Lambda}$ non-degenerate,
the CP-basis is unique up to a discrete subgroup of $SO(3)_H$; for the remaining
cases there is a continuous infinite of possible CP-basis, when one exists.
As for the CP-reflection direction, only when $\tilde{\Lambda}$ non-degenerate
and $\mathbf{M}\times\bs{\Lambda}_0\neq 0$ or
$\mathbf{M}\times\tilde{\Lambda}\mathbf{M}\neq 0$
($\mathbf{M}\parallel \bs{\Lambda}_0$) the direction is unique;
for $\mathbf{M}\parallel \bs{\Lambda}_0$ and
$\mathbf{M}\parallel\tilde{\Lambda}\mathbf{M}$ ($\tilde{\Lambda}$
non-degenerate), there are two possible directions.

At last, all higher order pseudoscalar invariants are either combinations of
lower order scalars or pseudoscalars, or is of the form Eq.\;\eqref{pseudoS} and
involves vectors with further applications of $\tilde{\Lambda}$, for example,
$\tilde{\Lambda}^n\mathbf{M}$; if the conditions (A) or (B) are valid, they all
remain in the principal plane defined by
$\{\mathbf{M},\bs{\mathbf{\Lambda}_0}\}$ or
$\{\mathbf{M},\tilde{\Lambda}\mathbf{M}\}$, which implies all
pseudoscalars of the form Eq.\;\eqref{pseudoS} are also null. This completes
(C).

Conditions (A) and (B) solve the problem of finding the minimum set of
reparameterization invariant conditions to test the CP-invariance of a
2HDM potential, a problem that was not completely solved in previous
approaches\;\cite{GH,branco:05,ivanov:05}.

For completeness, we compare the invariants of Eq.\;\eqref{Is:N=2} with that
of Ref.\,\onlinecite{GH} and arrive at the equalities
\eqarr{
\label{IY3Z}
I_{Y3Z}&=&I_2
\,,\\
I_{2Y2Z}&=& -\mn{\frac{1}{2}}I_1
\,,\\
I_{6Z}&=& -2I_{\Lambda_0}
\label{I6Z}
\,,\\
\label{I3Y3Z}
I_{3Y3Z}&=&\mn{\frac{1}{4}}[I_{M}+(\bM\!\cdot\!\bs{\Lambda_0})I_1
-M^2_0 I_2]
\,.
}
For Eqs.\,\eqref{IY3Z}--\eqref{I6Z}, the proportionality is assured by
dimensional counting, since these are the lowest order invariants by $SU(2)_H$
but not invariant by the corresponding CP-type transformation. The
proportionality constant can be found by restricting to particular values, for
example, $\lambda_6=-\lambda_7$\;\cite{GH}.
For Eq.\,\eqref{I3Y3Z}, the full calculation is necessary.
Also, the statement in Ref.\,\onlinecite{GH} that it is always possible to find
a basis when $\lambda_6=-\lambda_7$ can be seen here as the possibility of
rotating $\bs{\Lambda_0}$ in Eq.\;\eqref{Lambda:N=2} to the 3-direction.
Another example is the special point $\lambda_1=\lambda_2$ and
$\lambda_7=-\lambda_6$, which corresponds to $\bs{\Lambda}_0=0$ and the
condition for CP invariance only imposes conditions on $\mathbf{M}$ and
$\tilde{\Lambda}$. However, from the perspective of the development of this
section, we see $\mathbf{M}=0$ is as special a point as $\bs{\Lambda}_0=0$ is.
Only if the theory is CP-violating and one wants to classify the
violation in soft or hard violation\;\cite{ginzburg}, the two cases are
different.

\subsection{Conditions for spontaneous CP violation}
\label{subsec:scpv:N=2}

This issue has already been investigated in Ref.\;\onlinecite{GH}
using as the minimal representation the fundamental representation of $SU(2)_H$.
We will work out, instead, the conditions for spontaneous CP violation in the
2HDM using the adjoint representation.

We already explored the conditions to have explicit CP violation in $V(\Phi)$,
Eq.\;\eqref{V:N=2}. In such case after EWSB, the CP violating property will
remain in the potential. On the other hand, if the potential is CP conserving,
after EWSB, the theory could become CP violating if the vacuum is not invariant
by the CP-type transformation of the original potential.
We will concentrate on this spontaneously broken CP case.

Given the potential $V(\bbA)$ in Eq.\;\eqref{V2:N=2}, the spontaneously broken
potential is given by shifting the fields
\eqarr{
\label{Phi+<>:N=2}
\Phi&\rightarrow& \Phi+\aver{\Phi}
\\
\label{A+<>:N=2}
\bbA_\mu &\rightarrow& \bbA_\mu+\aver{\bbA_\mu}+\bbB_\mu
~,
}
where
\eqarr{
\label{<A>}
\aver{\bbA_ \mu}&\equiv&\mn{\frac{1}{2}} \aver{\Phi}^\dag\sigma_\mu\aver{\Phi}
\\
\bbB_ \mu&\equiv&\mn{\frac{1}{2}} \aver{\Phi}^\dag\sigma_\mu\Phi+
\mn{\frac{1}{2}} \Phi^\dag\sigma_\mu\aver{\Phi}
~.
}
The vacuum expectation values (VEVs), invariant by the $U(1)_{EM}$, can be
parametrized by
\eq{
\label{vev:N=2}
\aver{\Phi}=
\begin{pmatrix}
\aver{\Phi_1}\cr \aver{\Phi_2}
\end{pmatrix}
=
\frac{v}{\sqrt{2}}
\begin{pmatrix}
 0 \cr \displaystyle\frac{v_1}{v} \cr 0 \cr \displaystyle\frac{v_2}{v}e^{i\xi}
\end{pmatrix}
~,
}
where $v=\sqrt{v_1^2+v_2^2}=246\,\rm GeV$.
The parameters of Eq.\;\eqref{vev:N=2} have to obey the minimization constraints
\eq{
\label{min:N=2}
\frac{\partial}{\partial \phi_a^{\mt{(0)}*}}{V(\Phi)}\Big|_{\Phi=\aver{\Phi}}=
\frac{\partial}{\partial \bbA_\mu}{V(\bbA)}
\frac{\partial \bbA_ \mu}{\partial \phi_a^{\mt{(0)}*}}\Big|_{\Phi=\aver{\Phi}}
=(M_{\mu}+2\Lambda_{\mu\nu}\aver{\bbA_\nu})\mn{\frac{1}{2}}
(\sigma_\mu)_{ab}\aver{\phi_b^{\mt{(0)}}}
=0
~.
}
For the charged component the condition is trivial $\aver{\phi_a^{\mt{(+)}}}=0$.
The nontrivial ($\aver{\phi_a^{\mt{(0)}}}\neq 0$) solution for
Eq.\;\eqref{min:N=2} is conditioned by the existence of solutions
$\aver{\bbA_\mu}\neq 0$ of
\eq{
\label{min:sol:N=2}
\det[(M_{\mu}+2\Lambda_{\mu\nu}\aver{\bbA_\nu})\sigma_\mu]
=(M_{0}+2\Lambda_{0\nu}\aver{\bbA_\nu})^2-
(M_{i}+2\Lambda_{i\nu}\aver{\bbA_\nu})^2
=0
~,
}
provided that $V(\aver{\Phi})<V(0)$ and $\aver{\Phi}$ corresponds to an absolute
minimum. When Eq.\;\eqref{vev:N=2} is used, the parameterization of
$\aver{\bbA_{\mu}}$ is
\eq{
\label{vev:A:N=2}
\aver{\bbA_{\mu}}=\frac{v^2}{2}(1,\bv)=\frac{v^2}{2}
(1,\sin\theta_v\cos\xi,\sin\theta_v\sin\xi,\cos\theta_v)
~,
}
where $\tan\frac{\theta_v}{2}=\frac{v_2}{v_1}$; Eq.\;\eqref{vev:A:N=2} is just
the projective map of the complex number $v_2e^{i\xi}/v_1$ to the unit sphere.
Notice that the connection of the parameter $\theta_v$ used here with the more
usual parameter $\beta$ used in the MSSM description\;\cite{carena,Haber.2} is
given by $\tan\frac{\theta_v}{2}=\tan\beta$.

In case the potential in Eq.\;\eqref{V:N=2} has a CP symmetry, it can be
written in the CP-basis (or the {\it real basis}\;\cite{GH}) in the form of
Eq.\;\eqref{V4:N=2}. The CP transformations are just Eqs.\;\eqref{CP:Phi:N=2}
and \eqref{CP:A:N=2}. The potential after EWSB can be written as
\eqarr{
\label{V:Phi+<>:N=2}
V(\Phi+\aver{\Phi})&=&
V(\bbA)+V(\aver{\bbA})+
\Lambda_{\mu\nu}\bbB_\mu\bbB_\nu+
2\Lambda_{\mu\nu}\bbA_\mu(\aver{\bbA_\nu}+\bbB_\nu)
~,
}
which, in the CP-basis, have $M_2=0$, $\Lambda_{02}=0$ and
$\tilde{\Lambda}=\diag(\{\tilde{\lambda}_i\})$. The condition
$(M_{\mu}+2\Lambda_{\mu\nu}\aver{\bbA_\nu})\bbB_\mu=0$, derived from
Eq.\;\eqref{min:N=2} was used. By construction, if $\aver{\Phi}$ also
transformed under CP as $\Phi$, the potential in Eq.\;\eqref{V:Phi+<>:N=2} would
be CP invariant. However the invariance of the vacuum under any symmetry implies
the VEVs have to be invariant under the CP transformation.
Looking into the details, if we apply the transformations of
Eqs.\;\eqref{CP:Phi:N=2} and \eqref{CP:A:N=2} into Eq.\;\eqref{V:Phi+<>:N=2},
since $V(\Phi^*+\aver{\Phi})=V(\Phi+\aver{\Phi}^*)$ for an initial CP
invariant potential, the potential remains CP invariant after EWSB if, and only
if, \eqarr{
\label{cond:vevA:N=2}
(I_2\aver{\bbA})_i&=&\aver{\bbA_i} \\
\label{cond:vevPhi:N=2}
\bbB_i(\aver{\Phi}^*)&=&\bbB_i(\aver{\Phi})
~.
}
Equation \eqref{cond:vevA:N=2} implies $\aver{\bbA_2}=0$ and from the
parameterization of Eq.\;\eqref{vev:A:N=2} it implies $\xi=0,\pi$.
Then Eq.\;\eqref{cond:vevPhi:N=2} is automatically satisfied with
$\aver{\Phi}^*=\aver{\Phi}$. Actually, any solution of the form
$\aver{\Phi}^*=e^{i\alpha}\aver{\Phi}$ satisfies
Eq.\;\eqref{cond:vevA:N=2} but not Eq.\;\eqref{cond:vevPhi:N=2}.
The parameterization of Eq.\;\eqref{vev:N=2}, however, automatically takes
into account Eq.\;\eqref{cond:vevPhi:N=2} when Eq.\;\eqref{cond:vevA:N=2} is
satisfied. Thus, using such parameterization the analysis can be carried out
exclusively in the adjoint representation.

In a general basis, the conditions on the CP-basis investigated so far can be
translated to the following condition: if $V(\Phi)$ is CP invariant and it has a
nontrivial minimum $\aver{\Phi}\neq 0$, $V(\Phi+\aver{\Phi})$ is CP invariant
if, and only if, $\aver{\bbA_i}$ is in the principal plane defined by
$\{\mathbf{M},\bs{\Lambda}_0,\tilde{\Lambda}\mathbf{M}\}$. The more specific
conditions for $\{\aver{\bbA_i}\}$ to be in the latter principal plane are:
\begin{itemize}
\item[a)] If $\mathbf{M}\times \bs{\Lambda}_0\neq 0$,
$I(\{\aver{\bbA_i}\},\mathbf{M},\bs{\Lambda}_0)=0$.

\item[b)] If $\mathbf{M}\parallel\bs{\Lambda}_0$ and
$\tilde{\Lambda}\mathbf{M}\times \mathbf{M}\neq 0$,
$I(\{\aver{\bbA_i}\},\mathbf{M},\tilde{\Lambda}\mathbf{M})=0$.

\item[c)] If $\mathbf{M}\parallel\bs{\Lambda}_0$ and
$\tilde{\Lambda}\mathbf{M}\parallel\mathbf{M}$,
$I(\{\aver{\bbA_i}\},\mathbf{M},\tilde{\Lambda}\{\aver{\bbA_i}\})=0$.
\end{itemize}
The CP-reflection directions for (a) and (b) are the same as in (A) and (B) of
sec.\,\ref{sec:N=2}.
For (c), if $\{\aver{\bbA_i}\}\parallel \mathbf{M}$ the CP-reflection
direction is an eigenvector of $\tilde{\Lambda}$ perpendicular to $\mathbf{M}$;
otherwise $\{\aver{\bbA_i}\}\times \mathbf{M}$ is an eigenvector of
$\tilde{\Lambda}$ and it is the CP-reflection direction.
In the CP-basis, $\aver{\Phi}$ is real.

\section{$N\ge 2$ Higgs-doublets}
\label{sec:N>2}

For $N\ge 2$ Higgs-doublets $\Phi_a$, $a=1,\dots, N$, transforming as (2,1)
under $SU(2)_L\otimes U(1)_Y$, the general gauge invariant potential can be
written as\;\cite{endnote0}
\eq{
\label{V:N>2}
V(\Phi)=
Y_{ab}\Phi^\dag_{a}\Phi_b
+Z_{(ab)(cd)}(\Phi_{a}^\dag\Phi_b)^*
(\Phi_{c}^\dag\Phi_d) ~,
}
where
\eq{
\label{Phi:N>2}
\Phi=
\begin{pmatrix}\Phi_1 \cr \Phi_2 \cr \vdots \cr \Phi_N\end{pmatrix}
~.
}
We define then the minimal SM gauge invariants
\eq{
\label{Aab:N>2}
A_{ab}\equiv \Phi_a^\dag\Phi_b~,
}
and define a column vector of length $N^2$ by the ordering
\linebreak $(ab)=(11), (12), (13),\ldots, (1N), (21), \ldots (NN)$,
\eq{
\label{A:N>2}
A\equiv \Phi^\dag\otimes\Phi
=
\begin{pmatrix}
A_{11} \cr A_{12} \cr \vdots \cr A_{1N}\cr A_{21}\cr \vdots\cr
A_{NN} \end{pmatrix}
~.
}
Additionally we denote the pair of indices as $(ab)\equiv \sigma$, running as
Eq.\;\eqref{A:N>2}, and define the operation of change of labelling
$\widehat{\sigma}=(ba)$, if $\sigma=(ab)$, in such a way that
if $A_\sigma=A_{ab}$, then $A_{\sigma}^*=A_{ba}=A_{\widehat{\sigma}}$.
With this notation the quartic part of $V(\Phi)$ can be written
\eq{
V(\Phi)\Big|_{\Phi^4}=
A^*_\sigma Z_{\sigma\sigma'}A_{\sigma'}
\equiv A^\dag ZA
~.
}
This parameterization constrains $Z$ to be hermitian $Z^\dag=Z$,
\eq{
\label{hermZ}
(Z_{(ab)(cd)})^*=Z_{(dc)(ba)}
~ \text{or}~~
Z_{\sigma_1\sigma_2}^*=Z_{\sigma_2\sigma_1}~.
}
At the same time, because of $A^*_{\sigma_1}A_{\sigma_2}=
A^*_{\widehat{\sigma_2}}A_{\widehat{\sigma_1}}$, $Z$ has the property
\eq{
\label{propZ}
Z_{\sigma_1\sigma_2}=Z_{\widehat{\sigma_2}\widehat{\sigma_1}}
~.
}
Thus, $Z$ is a $N^2\times N^2$ hermitian matrix with the additional property of
Eq.\;\eqref{propZ}.
To count the number of independent variables of $Z$ we have to divide its
(complex)  elements into four sets: (d1) $N$ diagonal ($\sigma_1=\sigma_2\equiv
\sigma$ and  $\sigma=\hat{\sigma}$)  and (d2) $N(N-1)$ diagonal
($\sigma_1=\sigma_2\equiv \sigma$  and $\sigma\neq \hat{\sigma}$)  real elements
because of the  Hermiticity condition \eqref{hermZ};  (o1) $N(N-1)$ off-diagonal
($\sigma_1\neq \sigma_2$ but $\sigma_1=\hat{\sigma_2}$)  and (o2)
$N^2(N^2-1)-N(N-1)$ off-diagonal  ($\sigma_1\neq \sigma_2$ and
$\sigma_1\neq\hat{\sigma_2}$)  complex but not  all  independent elements. The
total is $N^4$ elements as it should be. The number  of independent real
parameters is then $N$ (d1) + $N(N-1)/2$ (d2) in the  diagonal real elements
[Eq.\;\eqref{propZ} only imposes conditions on the  elements in (d2)] and
$N(N-1)$ (o1) + $\frac{1}{2}[N^2(N^2-1)-N(N-1)]$ (o2) in  the off-diagonal
complex elements [Eq.\;\eqref{propZ} only imposes conditions  on the elements in
(o2)] summing up to $N^2(N^2+1)/2$.
For example, for $N=2$ there were $4(4+1)/2=10$ real
parameters corresponding to the real and complex parameters,
$\{\lambda_1,\lambda_2,\lambda_3,\lambda_4\}$ and
$\{\lambda_5,\lambda_6,\lambda_7\}$, respectively.

The horizontal transformation group is now $G=SU(N)_H$; a global phase can be
absorbed by $U(1)_Y$ symmetry as in the $N=2$ case.
We can define new variables, equivalent to Eq.\;\eqref{A:N=2}, as
\eq{
\label{bbA:N>2}
\bbA_\mu\equiv \mn{\frac{1}{2}}\Phi^\dag \lambda_\mu\Phi~,
~~\mu=0,1,\ldots,d,
}
where $\lambda_0=\sqrt{\frac{2}{N}}\,\id$ and $\{\lambda_i\}$ are the
$d={\rm dim\,}SU(N)=N^2-1$ hermitian generators of $SU(N)_H$ in the fundamental
representation obeying the normalization $\Tr[\lambda_i\lambda_j]=2\delta_{ij}$,
such that $\Tr[\lambda_\mu\lambda_\nu]=2\delta_{\mu\nu}$.
The new second order variables $\bbA_\mu$ transform as $\bar{N}\otimes N=d\oplus
{\bf 1}$, where $d$ denotes the adjoint representation. The index $\mu=0$
corresponds to the singlet component while the indices $\mu=i=1,...,d$
correspond to the adjoint, transforming under $SU(N)_H$ as
\eq{
\label{horizontal:R}
\bbA_i \rightarrow R_{ij}\bbA_j~.
}
The matrix $R_{ij}$ can be obtained from the fundamental represention $U$,
acting on $\Phi$ as Eq.\;\eqref{horizontal:N=2}, from the relation
\eq{
\label{Rij:U}
R_{ij}(U)=
\mn{\frac{1}{2}}\Tr[U^\dag \lambda_iU\lambda_j]
~.
}
If $U=\exp(i\bs{\theta}\cdot\bs{\lambda}/2)$,
\eq{
\label{Rij:theta}
R_{ij}(\bs{\theta})=
\exp[i\theta_kT_k]_{ij}~,
}
where $(iT_k)_{ij}=f_{kij}$. The coefficients $f_{ijk}$, are the structure
constants of $SU(N)_H$ defined by
\eq{
\label{fijk:sun}
[T_i,T_j]=if_{ijk}T_k~,
}
for any $\{T_i\}$ spanning the compact $SU(N)_H$ algebra $\cG$.
In particular, Eq.\;\eqref{fijk:sun} is valid for $\{\lambda_i/2\}$,
the fundamental representation generators. Since the structure constants of
Eq.\;\eqref{fijk:sun} are real, the adjoint representation of
Eqs.\;\eqref{Rij:U} and \eqref{Rij:theta} is real and thus it represents a
subgroup of $SO(d)$. It is only for $N=2$ the adjoint representation is the
orthogonal group itself.

The transformation matrix from $A_\sigma$~\eqref{A:N>2} to
$\bbA_\mu$~\eqref{bbA:N>2} can be obtained from the completeness relation of
$\{\lambda_\mu\}$\;\cite{fierz}:
\eq{
\label{complete}
\mn{\frac{1}{2}}(\lambda_\mu)_{ab}(\lambda_\mu)_{cd}=
\delta_{ad}\delta_{cb}~.
}
In the notation where $\sigma_1=(ab)$ and $\sigma_2=(cd)$, we can write
Eq.\;\eqref{complete} in the form
\eq{
\label{complete2}
2C_{\mu\sigma_1}C_{\mu\sigma_2}=
\delta_{\sigma_1\widehat{\sigma_2}}~,
}
where
\eq{
\label{C}
C_{\mu\sigma_1}\equiv\mn{\frac{1}{2}}(\lambda_{\mu})_{\sigma_1}~.
}
Equation \eqref{complete2} implies $C^{-1}_{\sigma\mu}\equiv
2C_{\mu\widehat{\sigma}}$, since the inverse is unique.
The definition of Eq.\;\eqref{C} enable us to write Eq.\;\eqref{bbA:N>2} in the
form
\eq{
\label{A->bbA:N>2}
\bbA_\mu=C_{\mu\sigma}A_{\sigma}~.
}

By using the inverse of Eq.\;\eqref{A->bbA:N>2} we can write the potential of
Eq.\;\eqref{V:N>2} in the same form of Eq.\;\eqref{V2:N=2},
\eqarr{
\label{V2:N>2}
V(\bbA)&=&M_\mu\bbA_\mu+\Lambda_{\mu\nu}\bbA_\mu\bbA_\nu~,
\\
\label{V3:N>2}
&=&M_0\bbA_0+\Lambda_{00}\bbA_0^2+
M_i\bbA_i+2\Lambda_{0i}\bbA_0\bbA_i+
\tilde{\Lambda}_{ij}\bbA_i\bbA_j~,
}
where
\eqarr{
M_\mu&\equiv& \Tr[Y\lambda_\mu]~,
\\
\Lambda_{\mu\nu}&\equiv&
C^{-1*}_{\sigma_1\mu}Z_{\sigma_1\sigma_2}C^{-1}_{\sigma_2\nu}~,
\\
\tilde{\Lambda}&=&\{\Lambda_{ij}\}~, ~~i,j=1,2\ldots,N.
}
Using the properties of Eqs.\;\eqref{hermZ} and \eqref{propZ} of $Z$, we can see
$\Lambda$ is a $N^2\times N^2$ real and symmetric matrix, hence with
$N^2(N^2+1)/2$ real parameters, the same number of parameters of $Z$.
The rank-2 tensor $\tilde{\Lambda}$ transforms under $G$ as $(d\otimes
d)_S$ and it forms a reducible representation. (See appendix \ref{dxd}.)

The procedure to find the CP-basis can be sought in some analogy with the
$N=2$  case. The difficulty for $N>2$, however, is that the existence of a
horizontal transformation on the vector $\bbA_i$, defined by
Eq.\;\eqref{horizontal:R}, capable of diagonalizing $\tilde{\Lambda}$, is not
always guaranteed and it depends on the form of $\tilde{\Lambda}$ itself.

Nevertheless, the diagonalization of $\tilde{\Lambda}$ is not strictly
necessary. To see this, we have to analyze the CP properties of $\Phi$ and
$\bbA_i$. Firstly, any CP-type transformation can be written as a combination of
a horizontal transformation and the canonical CP-transformations
\eqarr{
\Phi(x) &\stackrel{CP}{\rightarrow}&  \Phi^*(\hat{x})~,
\\
\bbA_0(x) &\stackrel{CP}{\rightarrow}&  \bbA_0(\hat{x})~,
\\
\label{reflection:N>2}
\bbA_i(x) &\stackrel{CP}{\rightarrow}& -\eta_{ij}\bbA_j(\hat{x})~.
}
Equation \eqref{reflection:N>2} represents the canonical CP-reflection defined
by the CP-reflection matrix $\eta$ given by
\eq{
\eta_{ij}= -\mn{\frac{1}{2}}\Tr[\lambda^{\tp}_i\lambda_j]~,
}
which means $\lambda_i^{\tp}= -\eta_{ij}\lambda_j$. The mapping
$\lambda_i \stackrel{\psi}{\rightarrow} -\lambda_i^{\tp}=\eta_{ij}\lambda_j$ is
the contragradient automorphism in the Lie algebra\;\cite{GR}, which I will
denote by $\psi$. Such automorphism maps the fundamental representation to
the antifundamental representation, $D(g)\rightarrow D^*(g)$ (these two
representations are not equivalent for $N> 2$). All the irreducible
representations (irreps) we are treating here [$d$ and all components
in $(d\otimes d)_S$] are self-conjugate\,\cite{slansky,GR} and, indeed, they are
real representations.

To set a convention, we will use the following ordering for the basis of the Lie
algebra $\cG$ of $G$:
\eq{
\label{ordering}
\{T_i\}=\{h_i, \cs_{\alpha}, \ca_\alpha\}.
}
The set $\{h_i\}$ spans the Cartan subalgebra (CSA) $\ft_r$ and the set
$\{\ca_\alpha\}$, denoted by $\ft_q$, are the generators of the real
$H=SO(N)$ subgroup of $G=SU(N)$. The remaining subspace spanned by
$\{\cs_{\alpha}\}$ will be denoted by $\tilde{\ft}_{q}$ and the sum
$\ft_r\oplus \tilde{\ft}_q\equiv \ft_p$ represents the generators of the coset
$G/H$. Notice that $\ft_p$ and $\ft_q$ are invariant by the action of the
subgroup $H$ and hence they form representation spaces for $H$ (see appendix
\ref{app:d=p+q}).
We will use the symbols $\{h_i,\cs_\alpha,\ca_\alpha\}$ to denote either the
abstract algebra in the Weyl-Cartan basis or the fundamental representation
of them.

The dimensions of these subspaces of $\cG$ are respectively,
$r=\mathrm{rank}G=N-1$, $q=(d-r)/2=N(N-1)/2$ and $p=(d+r)/2=N(N+1)/2-1$; $q$
denotes the number of positive roots in the algebra and $\alpha$ are the
positive roots that label the generators \eqarr{
\label{cs}
\cs_\alpha&=&\frac{E_{\alpha}+E_{-\alpha}}{2} ~,
\\
\label{ca}
\ca_\alpha&=&\frac{E_{\alpha}-E_{-\alpha}}{2i} ~.
}
The $E_{\alpha}$ are the ``ladder'' generators in the Cartan-Weyl basis.
For example, for the fundamental representation of $SU(3)$, we have in terms of
the Gell-Mann matrices\;\cite{IZ}, $\{h_i\}=\{\lambda_3/2,\lambda_8/2\}$,
$\{\cs_\alpha\}=\{\lambda_1/2, \lambda_6/2, \lambda_4/2 \}$ and
$\{\ca_\alpha\}=\{\lambda_2/2, \lambda_7/2, \lambda_5/2 \}$.
The two last subspaces are ordered according to $\alpha_1,\alpha_2$ and
$\alpha_3=\alpha_1+\alpha_2$.
Notice that in such representation $\cs_{\alpha}$ are symmetric matrices and
$\ca_\alpha$ are antisymmetric matrices.

With such ordering the CP-reflection matrix is
\eq{
\eta=\begin{pmatrix}
 -\id_p & 0 \cr 0 & \id_q
\end{pmatrix}
~.
}
Thus, we see that the application of the automorphism $\psi$ separates $\cG$
into an odd part $\ft_p$ and an even part $\ft_q$, which constitutes a
subalgebra\;\cite{Gilmore}.
The condition for CP-invariance of the term containing $\tilde{\Lambda}$ in
Eq.\;\eqref{V3:N>2} is then the existence of a group element $g$ such that
\eq{
\label{cond1:lamb}
\text{\bf cond.\,1:}\quad
\eta R(g)\tilde{\Lambda}R(g^{-1})\eta=R(g)\tilde{\Lambda}R(g^{-1})~,
}
where $R(g)\equiv R(U)$ is an element in the adjoint
representation of $SU(N)$~\eqref{Rij:U}. Equation \eqref{cond1:lamb} is
equivalent, in this representation, to the statement: exists a $R(U)$ in the
adjoint representation of SU(N), such that $R\tilde{\Lambda}R^{^{\tp}}$ is
block diagonal $p\times p$ superior and $q\times q$ inferior. Thus full
diagonalization is not necessary.

Now suppose a $g$ satisfying {\bf cond.\,1} exists\;\cite{endnote2}. We
can write Eq.\;\eqref{V3:N>2} in the basis defined by one representative $g$,
\eq{
V(\bbA)=M_0\bbA_0+\Lambda_{00}\bbA_0^2+
M'_i\bbA'_i+2\Lambda'_{0i}\bbA_0\bbA'_i+
\tilde{\Lambda}'_{ij}\bbA'_i\bbA'_j~,
}
where $M'_i=R(g)_{ij}M_j$ and  $\Lambda'_{0i}=R(g)_{ij}\Lambda_{0j}$.
The necessary conditions for $V(\bbA)$ to be CP invariant are
\eqarr{
\label{cond2:M}
\textbf{cond.\,2a:}\quad
M'_i&=&-\eta_{ij}M'_j \\
\label{cond2:lamb0}
\textbf{cond.\,2b:}\quad
\Lambda'_{0i}&=&-\eta_{ij}\Lambda'_{0j}
~.
}
Of course, there can be more than one distinct coset satisfying {\bf cond.\,1},
and, then, conditions \eqref{cond2:M} and \eqref{cond2:lamb0} have to be
checked for all these cosets. If for every coset satisfying {\bf cond.\,1},
there is no coset satisfying {\bf cond.\,2a} and {\bf cond.\,2b}, then the
potential is CP violating. Otherwise, $g$ satisfying {\bf conds.\,1},
{\bf 2a} and {\bf 2b} defines a CP-basis and the potential is CP
invariant.

Let us analyze further the {\bf conds.\,1}, {\bf 2a} and {\bf 2b}. To do that,
we denote by $V=\mathbb{R}^d\sim \mathcal{G}$ the adjoint representation space,
isomorphic to the algebra vector space. The automorphism $\psi$ separates the
space $V$ into two subspaces $V=V_p\oplus V_q$, one odd ($V_p$) and one even
($V_q$) under the automorphism: \eqarr{
\label{etaVpq}
\eta \mathbf{v}&=&-\mathbf{v}, ~~\text{if $v\in V_p$},\cr
\eta \mathbf{v}&=&\phantom{-}\mathbf{v}, ~~\text{if $v\in V_q$}.
}
They correspond respectively to $\ft_p$ and
$\ft_q$, subspaces of $\mathcal{G}=\ft_p\oplus \ft_q$.
The correspondence between $\cG$ and $V$ is given by Eq.\;\eqref{x->tx}.
With this notation, considering the matrix $\tilde{\Lambda}$ is a linear
transformation over $V$, {\bf conds.} $\bf 1, 2.a$ and $\bf 2.b$ imply
that there should be two subspaces $V'_p$ and $V'_q$ of $V=V'_p \oplus V'_q$
invariant by $\tilde{\Lambda}$ and both $\mathbf{M}$ and $\bs{\Lambda}_{0}$
should be in $V'_p$. Moreover, the two subspaces should be connected to $V_p$
and $V_q$ by a group transformation, i.e.,  $V'_p=R(g)V_p$ and
$V'_q=R(g)V_q$ for some $g$ in $G$.

The explicit search for the matrices satisfying {\bf conds.} $\bf 1$, $\bf
2.a$ and $\bf 2.b$ is a difficult task. We can seek, instead, invariant
conditions based on group invariants, analogously with what was done to the
$N=2$ case. For that purpose, it will be shown in the following that generalized
pseudoscalar invariants\;\cite{endnote3}, analogous to the true pseudoscalars
\eqref{pseudoS}, can still be constructed with respect do
$SU(N)$ and any such quantity should be zero for a CP-invariant potential.

The generalized pseudoscalar is defined as a trilinear totally antisymmetric
function of vectors in the adjoint, defined by
\eq{
\label{I:N>2}
I(\mathbf{v}_1,\mathbf{v}_2,\mathbf{v}_3)\equiv
f_{ijk}v_{1i}v_{2i}v_{3i}~.
}
We keep the same notation as in Eq.\;\eqref{pseudoS}, noticing that
Eq.\;\eqref{I:N>2} corresponds to a more general case. We
can also define the analogous of the vector product in three dimensions, as
\eq{
\label{prod:a}
(\mathbf{v}_1\wedge \mathbf{v}_2)_i \equiv
f_{ijk}v_{1j}v_{2k}~.
}
From Eq.\;\eqref{fijk} and $\lambda^{\tp}_i = -\eta_{ij}\lambda_j$ we see that
\eq{
\label{fijk:psi}
f_{ijk}\eta_{ia}\eta_{jb}\eta_{kc}=f_{abc}~,
}
which means $\psi$ indeed represents an automorphism in the algebra.
However, the CP-reflection of Eq.\;\eqref{reflection:N>2} acts with the opposite
sign compared to the automorphism $\psi$.
Therefore, the quantity in Eq.\;\eqref{I:N>2} is invariant by $SU(N)_H$
transformations but changes sign under a CP transformation, i.e., a
CP-reflection. Thus we see the trilinear function \eqref{I:N>2} behaves as a 
pseudoscalar under a CP-reflection. Such  property means that any pseudoscalar
of the form Eq.\;\eqref{I:N>2}, constructed with the parameters of a
CP-invariant potential $V(\bbA)$ should be zero.

The pseudoscalar invariants of lowest order, constructed with
$\{\mathbf{M},\bs{\Lambda}_0,\tilde{\Lambda}\}$, are of the same form as
in Eqs.\;\eqref{Is:N=2}. We will see, however, that the vanishing of these
quantities may not guarantee the CP-invariance of $V(\bbA)$.
Let us exploit further the properties of the CP-reflection
\eqref{reflection:N>2}. For that end, it is known that an additional trilinear
scalar, which is totally symmetric, can be defined for $N>2$ as
\eq{
\label{J} J(\mathbf{v}_1,\mathbf{v}_2,\mathbf{v}_3)\equiv
d_{ijk}v_{1i}v_{2j}v_{3k}~,
}
as well as a ``symmetric'' vector product
\eq{
\label{prod:s}
(\mathbf{v}_1\vee \mathbf{v}_2)_i \equiv
d_{ijk}v_{1j}v_{2k}~.
}
The coefficient $d_{ijk}$ is the totally symmetric 3-rank tensor of $SU(N)$
defined by Eq.\;\eqref{anticom:F}.
The behavior of the scalar $J$ \eqref{J} is opposite to the scalar $I$
\eqref{I:N>2}, since it changes sign under the contragradient automorphism, as
can be seen by Eq.\;\eqref{dijk}, and remain invariant under  a CP-type
reflection.

Using the two trilinear invariants $I$ and $J$, the following relations can
be obtained for any $V'_p=R(g)V_p$ and $V'_q=R(g)V_q$:
\eqarr{
\label{IJ:Vpq}
I(V'_p,V'_p,V'_p)&=&0~,\cr
I(V'_q,V'_q,V'_p)&=&0~,\cr
J(V'_q,V'_q,V'_q)&=&0~,\cr
J(V'_p,V'_p,V'_q)&=&0~.
}
These relations can be proved by noting that they are invariant under
transformations in $G$ and it can be evaluated with the corresponding vectors in
the original subspaces $V_p$ and $V_q$. Using Eqs.\;\eqref{fijk:psi} (for
$d_{ijk}$ the opposite sign is valid) and \eqref{etaVpq}, the invariants in
Eq.\;\eqref{IJ:Vpq} are equal to their opposites, which imply they are null. For
example, $f_{ijk}v_{1i}v_{2j}v_{3k}=
f_{ijk}\eta_{ia}\eta_{jb}\eta_{kc}v_{1a}v_{2b}v_{3c}=
-f_{ijk}v_{1i}v_{2j}v_{3k}=0$ for $\mathbf{v}_1,\mathbf{v}_2,\mathbf{v}_3$ in
$V_p$. (The relations \eqref{IJ:Vpq} can also be proved by using the explicit
representations for $f_{ijk}$ and $d_{ijk}$, appendix \ref{f,d:ijk}.)
Moreover, the relations in Eq.\;\eqref{IJ:Vpq} imply
\eqarr{
V'_p\wedge V'_p &\subset& V'_q ~,\cr
V'_p\wedge V'_q &\subset& V'_p ~,\cr
V'_q\wedge V'_q &\subset& V'_q ~,
}
and
\eqarr{
V'_p\vee V'_p &\subset& V'_p ~,\cr
V'_p\vee V'_q &\subset& V'_q ~,\cr
V'_q\vee V'_q &\subset& V'_p ~,
}
since the choice of vectors in each subspace is arbitrary and the two subspaces
are disjoint and covers the whole vector space $V$.
The first relation in Eq.\;\eqref{IJ:Vpq} confirms that if {\bf conds.}
$\bf 1,2.a$ and $\bf 2.b$ are satisfied, all $I$ invariants are indeed null.

Let us now analyze {\bf cond.\,1}. If {\bf cond.\,1} is true, $\tilde{\Lambda}$
should have two invariant subspaces $V'_p$ and $V'_q$ connected to $V_p$ and
$V_q$ by the same group element. Since $\tilde{\Lambda}$ is real and symmetric,
it can be diagonalized by $SO(d)$ transformations with real eigenvalues. The $d$
orthonormal eigenvectors of $\tilde{\Lambda}$, denoted by $e'_i$,
$i=1,2,\ldots,d$, form a basis for $V$ (if $\tilde{\Lambda}$ is degenerate, find
orthogonal vectors in the degenerate subspace). Any set of eigenvectors spans an
invariant subspace of $\tilde{\Lambda}$. $q$ of them should span a subspace
connected to $V_q$ while the remaining $p$ eigenvectors should span the
orthogonal complementary subspace connected to $V_p$. There is a criterion to
check if a given vector $\bf v$ is in some $V'_q$. For $q$ vectors, additional
criteria exist to check if they form a vector space and if they are closed under
the algebra $\cG$. These criteria follow from the fact that $V_q$ is isomorphic
to $\ft_q$ which forms a subspace of the ($i$ times) $N\times N$ real
antisymmetric matrices. Any antisymmetric matrix $M$ has $\Tr[M^{2k+1}]=0$. The
converse is true in the sense that for $M$ hermitian, $\Tr[M^{2k+1}]=0$ for all
$2k+1\le N$ imply $M$ can be conjugated by $SU(N)$ to an antisymmetric matrix,
i.e., in $\ft_q$. (See appendix \ref{app:A} for the proof.)

Therefore, $\bf v$ belongs to a $V'_q$ if, and only if,
\eq{
\label{v->Vq}
\mn{\frac{1}{2}}\Tr[(\mathbf{v}\cdot\bs{\lambda})^{2k+1}]
=J_{2k+1}(\mathbf{v},\mathbf{v},\ldots,\mathbf{v})
=0~,~~\text{for all $2k+1\le N$}.
}
We have introduced the $n$-linear symmetric function
\eq{
J_{n}(\mathbf{v}_1,\mathbf{v}_2,\ldots,\mathbf{v}_n)
\equiv
\Gamma^{\mt{(n)}}_{i_1 i_2\cdots i_n}
v_{1\,i_1}v_{2\,i_2},\ldots,v_{n\,i_n}
~,
}
which depends on the rank-$n$ totally symmetric tensor
\eq{
\label{Gamma}
\Gamma^{\mt{(n)}}_{i_1 i_2\cdots i_n}\equiv
\frac{1}{n!}\sum_{\sigma}
\mn{\frac{1}{2}}\Tr[\lambda_{\sigma(i_1)}\lambda_{\sigma(i_2)}\ldots
\lambda_{\sigma(i_n)}] ~,~~n\ge 2~,
}
where $\sigma$ denotes permutations among $n$ elements and the sum runs over
all possible permutations.
The tensor in Eq.\;\eqref{Gamma} are the tensors used to construct the $r$
Casimir invariants of any representation of $SU(N)$\;\cite{ORaif}.
In particular, $\Gamma^{\mt{(3)}}_{ijk}=d_{ijk}$ and $J_3=J$.

For two vectors $\mathbf{v}_1$ and $\mathbf{v}_2$, each one satisfying
Eq.\;\eqref{v->Vq}, the linear combination $c_1 \mathbf{v}_1+
c_2\mathbf{v}_2$ is also in some $V'_q$ if, and only if,
\eqarr{
\label{v1+v2}
\mn{\frac{1}{2}}\Tr[(c_1
\mathbf{v}_1\cdot\bs{\lambda}+ c_2\mathbf{v}_2\cdot\bs{\lambda})^{2k+1}]
=0~, ~~\text{for all $2k+1\le N$.}
}
In general,
\eqarr{
\mn{\frac{1}{2}}\Tr[(c_1
\mathbf{v}_1\cdot\bs{\lambda}+ c_2\mathbf{v}_2\cdot\bs{\lambda})^{n}]
&=&\sum_{m=0}^{n}\binom{m}{n}
c_1^m c_2^{n-m}
J_n(\underbrace{\mathbf{v}_1,\mathbf{v}_1,\ldots,\mathbf{v}_1},
\underbrace{\mathbf{v}_2,\ldots,\mathbf{v}_2})
~.
\cr&&\phantom{A_{m=0}^{n}(m)c_1^m c_2^{n-m}
J_n(\mathbf{v}_1,\mathbf{v}_1}m \hs{3.9em} n-m
}
Since the coefficients $c_1$ and $c_2$ are arbitrary, Eq.\;\eqref{v1+v2},
requires
\eq{
\label{v1+v2:J}
J_{2k+1}(\mathbf{v}_1,\ldots,\mathbf{v}_1,\mathbf{v}_2,\ldots,\mathbf{v}_2)
=0
}
for $2k+1\le N$ and all combinations of $\mathbf{v}_1$ and $\mathbf{v}_2$.

The generalization for a set of $q$ normalized eigenvectors of
$\tilde{\Lambda}$, labelled as $e'_{p+i}$, $i=1,\ldots,q$, is straightforward.
They form a $q$-dimensional vector space $V'_q$ if, and only if, each vector
satisfy Eq.\;\eqref{v->Vq} and any combination of $m\le q$ vectors satisfy
Eq.\;\eqref{v1+v2:J}. To guarantee that they are closed under the algebra $\cG$,
compute the $I$ invariants using any two vectors in $V'_q$ and one vector in
$V'_p$, as in the second relation of Eq.\;\eqref{IJ:Vpq}: they should all be
null. This conditions attest that the vector space $V'_q$ is $q$-dimensional and
forms a subalgebra of  $\cG$. That the subalgebra isomorphic to $V'_q$ is
semisimple and compact can be checked by Cartan's criterion: the Cartan metric,
as in Eq.\;\eqref{norm:fijk}, have to be positive definite\;\cite{Gilmore}.  It
remains to be checked if $V'_q$  is indeed connected to $V_q$ by a group
element.

At this point, we can see an example for which the vanishing of the
$I$-invariants \eqref{Is:N=2} [generalized to $N>2$ using Eq.\;\eqref{I:N>2}]
does not guarantee the CP-invariance of the potential:
If $\mathbf{M}$ and $\bs{\Lambda}_0$ are orthogonal eigenvectors of
$\tilde{\Lambda}$, all $I$-invariant are null, but nothing can be said about
$\tilde{\Lambda}$ satisfying {\bf cond.\,1}. Even if a $V'_q$ can be found using
the procedure above, if it contains at least one of $\mathbf{M}$ or
$\bs{\Lambda}_0$, the potential is CP violating.

For $N=3$, the problem of finding the necessary and sufficient conditions for
CP-invariance can be completely solved. In this case, the only
nontrivial symmetric function is the $J$ invariant in Eq.\;\eqref{J}. The
numerology is $d=8$, $r=2$, $q=3$ and $p=5$ and thus, $V'_q$ is three
dimensional. It can be proved\;\cite{michel} that any three dimensional
subalgebra is either conjugated to the $SU(2)$ subalgebra spanned by
$\{\lambda_1/2,\lambda_2/2,\lambda_3/2\}$ or to the real $SO(3)$ subalgebra
spanned by $\{\lambda_2/5,\lambda_7/2,\lambda_5/2\}$, in Gell-Mann's notation.
If three eigenvectors $e'_6,e'_7,e'_8$ of $\tilde{\Lambda}$ satisfy conditions
\eqref{v->Vq} and \eqref{v1+v2:J}, an additional condition to distinguish
between the two equidimensional subalgebras is to use\;\cite{michel}
\eq{
\label{cond:N=3}
|I(e'_6,e'_7,e'_8)|=\frac{1}{2}~,
}
which is satisfied only if the subalgebra is conjugate to $SO(3)$. For the
subalgebra conjugate to $SU(2)$ the value for Eq.\;\eqref{cond:N=3} is unity.
This fact can be understood by observing that
$\{\lambda_2/5,\lambda_7/2,\lambda_5/2\}$ are half of the usual generators of
$SO(3)$ in the defining representation, giving for the structure constants
restricted to the subalgebra,
$I(e'_{p+i},e'_{p+j},e'_{p+k})=\frac{1}{2}\varepsilon_{ijk}$, $i,j,k=1,2,3$,
after choosing appropriately the ordering for those three vectors.
In addition, if $\bf cond.\,1$ is true, an appropriate basis for the
$V(\bbA)$ can be chosen to be the one with $q\times q$ inferior block of
$\tilde{\Lambda}$ diagonal. Such choice is possible because when
$\tilde{\Lambda}$ is block diagonal $p\times p$ and $q\times q$, a
transformation in $SO(3) \subset SU(3)$ can still make the inferior block
diagonal. For $N>3$ that procedure is no longer guaranteed since the $q\times q$
blocks is transformed by  the adjoint representation of $SO(N)$, which differs
from the defining representation (see appendix \ref{app:d=p+q}).

Once the vector space $V'_q$ is found, when such space is unique up to
multiplication by the subgroup $SO(3)$, the vectors $\mathbf{M}$ and
$\bs{\Lambda}_0$ should be in the orthogonal subspace $V'_p$, i.e.,
$\mathbf{M}\cdot e'_{p+i}=0$ and $\bs{\Lambda}_0\cdot e'_{p+i}=0$ for all
$i=1,2,3$. Otherwise the potential $V(\bbA)$ is CP-violating.

\subsection{Conditions for spontaneous CP violation}
\label{subsec:scpv:N>2}

Let us briefly analyze the conditions for spontaneous CP-violation for a
potential $V(\Phi)$\,\eqref{V2:N>2} that is CP-invariant before EWSB.

The analysis can be performed in complete analogy with
Sec.\;\ref{subsec:scpv:N=2}. The minimization equation in this case are
identical to Eqs.\;\eqref{min:N=2}, after replacing the matrices $\sigma_\mu$ by
the corresponding $\lambda_\mu$ in Eq.\;\eqref{bbA:N>2}. The same replacement
applies to the first member of Eq.\;\eqref{min:sol:N=2}.
The conditions \eqref{cond:vevA:N=2} and \eqref{cond:vevPhi:N=2} in the CP-basis
are replaced by
\eqarr{
\label{scpv:Phi:N>2}
\aver{\Phi}^* &=&\aver{\Phi} ~,\\
\label{scpv:A:N>2}
\aver{\bbA_i} &=&-\eta_{ij}\aver{\bbA_j}~.
}
A suitable generalization of parameterization \eqref{vev:N=2} for
$N>2$ can be defined as
\eq{
\label{<Phi>:N>2}
\aver{\Phi}=
\frac{v}{\sqrt{2}}\,(U_v\, e_N)\otimes e_2~,
}
where $e_N=(0,0,\ldots,0,1)^{\tp}\in \mathbb{C}^N$, $e_2=(0,1)^{\tp}$ is the
$SU(2)_L$ breaking direction and $U_v$ is a $SU(N)_H$ transformation.
The parameterization \eqref{<Phi>:N>2} is justified because any vector
$z=(z_1,z_2,\ldots,z_N)^{\tp}$ in $\mathbb{C}^N$ can be transformed by a $SU(N)$
transformation into $z'=(0,0,\ldots,|z|)^{\tp}$\;\cite{ORaif}.
If Eq.\;\eqref{scpv:Phi:N>2} is true $U_v$ is real and belongs to the
real subgroup $SO(N)$.

In a general basis, it is necessary that $\{\aver{\bbA_i}\} \in V'_p$, i.e.,
the vector corresponding to the VEV have to be in the same subspace as
$\mathbf{M}$ and $\bs{\Lambda}_{0}$, which is true if
$\{\aver{\bbA_j}\}\cdot e'_{p+i}=0$, $i=1,\ldots,q$, for $\{e'_{p+i}\}$
spanning the subspace $V'_q$ invariant by $\tilde{\Lambda}$.
For $N=3$, such conditions are sufficient to guarantee that $V(\Phi
+\aver{\Phi})$ is also CP-invariant.

\section{Conclusions and Discussions}
\label{sec:concl}

The NHDMs are simple extensions of the SM for which the presence of a
horizontal space allows the possibility of ``rotating'' the basis in such space
without modifying the physical content of the theory, {\it e.g.}, CP symmetry or
asymmetry. For $N$ similar SM Higgs-doublets, which are complex, the
relevant reparameterization transformations form a $SU(N)$ group.
Restricted to the scalar potential sector, due to the rather restricted bilinear
form of the minimal gauge invariants, the NHDM potential can be written in terms
of the adjoint  representation of $SU(N)_H$.
The CP-type transformations act as ``reflections'', the CP-reflections, on the
parameters written as vectors and tensors of the adjoint.
Therefore, the scalar potential of the NHDMs are CP-invariant if, and
only if, one can find a CP-reflection that leaves the potential invariant.
In addition, the analysis in the adjoint representation was shown to be much
easier to carry out than the tensor analysis based on the fundamental and
antifundamental representations. Of course if other representations
that can not be written in terms of the adjoint are present, the analysis
invariably would require the fundamental representations. For example, to extend
this analysis to the Yukawa sector of the NHDMs, the fundamental representation
is necessary there.

For $N=2$, with the fortunate coincidence of the adjoint of $SU(2)$ being the
rotation group in three dimensions, the full analysis is facilitated by the
possible geometrical description. All the necessary and sufficient conditions
for CP violation can be formulated for the 2HDM scalar potential sector. Those
conditions can be formulated in terms of basis invariants which coincided with
previously found ones\;\cite{GH}, except for proportionality constants.
(A comparison between the invariants in Refs.\;\onlinecite{GH} and
\onlinecite{branco:05} is given in Ref.\;\onlinecite{ivanov:05}.) For
CP-invariant potentials, this method  also enabled us to find the explicit CP
transformation in any basis and the procedure to reach the real basis. For
CP-violating potentials, the canonical form of Eq.\;\eqref{V4:N=2} still defines
a standard form, besides the physical Higgs-basis\;\cite{lavoura:94,davidson},
to compare among the various 2HDMs: two 2HDM potentials are physically
equivalent if they have the same form in the canonical CP-basis. (For
convention, use the basis for which the eigenvalues of $\tilde{\Lambda}$ is in
decreasing order.) This CP-basis also makes the soft/hard classification of
CP-violation\;\cite{ginzburg} easier to perform: From Eq.\;\eqref{V4:N=2}, we
see the potential $V(\bbA)$ violates the CP-symmetry hardly only if the fourth
term is CP violating, i.e., if $I_{\Lambda_0}$ \eqref{IL0} is not null;
otherwise, the potential has soft CP violation through the third term or it is
CP symmetric. From Eq.\;\eqref{V:Phi+<>:N=2}, we see the spontaneous CP
violation only occurs softly.

For $N=3$, the necessary and sufficient conditions for CP-violation can still be
formulated in a systematic way. However, these conditions may possibly be
reduced to fewer and more strict conditions. Such reduction requires a more
detailed study of the relation between the invariants \eqref{Is:N=2}--\eqref{I2}
and the described procedure to check the CP symmetry or asymmetry. In case the
potential is CP-invariant, the explicit procedure to reach a real basis (among
infinitely many) is also lacking in this context and for $N>3$ as well.

For $N>3$, necessary conditions for CP-invariance in the NHDM potential can be
found but whether those conditions are sufficient or can be supplemented to
be sufficient is an open question. The answer lies in the classification and
perhaps parameterization of the orbital structure of the adjoint representation
of the $SU(N)$ group. In any case, if a result similar to $N=3$ can be found,
i.e., if any $SO(N)$ subalgebra of the $SU(N)$ algebra is conjugated to the real
$SO(N)$ subalgebra, the problem is practically solved.

Another possible approach would be the study of the automorphism properties of
the irreducible representation of $SU(N)$ contained in $\tilde{\Lambda}$
that are larger than the adjoint. For example, $\tilde{\Lambda}$ for $N\ge 3$
contains a component transforming under the adjoint representation
(see table \ref{tab:sun} in appendix \ref{dxd} and appendix \ref{app:adj:S}).
For this component it always exists a transformation capable of transforming it
to satisfy Eq.\;\eqref{cond1:lamb}. For higher dimensional irreps a detailed study
is not known to the author.

To conclude, the method presented here illustrates that using the
adjoint representation as the minimal nontrivial representation can have
substantial advantage over the fundamental representation treatmens to handle
the freedom of change of basis within a large horizontal space.
Inherent to that was the notion of CP-type transformations as automorphisms in
the group of horizontal transformations. Such notion was useful to distinguish
the CP invariance/violation (explicit/spontaneous) properties of the theory and
to construct the CP-odd basis invariants.

\appendix
\section{Notation and conventions}
\label{app:conv}

We use for the fundamental representation of $SU(N)$ the $N\times N$
traceless hermitian matrices $\{F_a\}\equiv\{\mn{\frac{1}{2}}\lambda_a\}$
normalized as
\eq{
\label{norm:F}
\Tr[F_iF_j]=\mn{\frac{1}{2}}\delta_{ij}~.
}
The number of generators is $d={\rm dim}SU(N)=N^2-1$.
The matrices $\lambda_a$ are generalizations of the Gell-Mann matrices for
$SU(3)$\;\cite{IZ}.

The compact semisimple Lie algebra is defined by
\eq{
\label{lie:F}
[F_i,F_j]=if_{ijk}F_k~,
}
which is satisfied for any represention $D(F_a)$.
By using the convention of Eq.\;\eqref{norm:F}, we have the relation
\eq{
\label{fijk}
f_{ijk}=\frac{2}{i}\Tr\big[[F_i,F_j]F_k\big]
=\frac{1}{4i}\Tr\big[[\lambda_i,\lambda_j]\lambda_k\big]
~.
}
The Cartan metric in the adjoint representation reads
\eq{
\label{norm:fijk}
\sum_{j,k=1}^{d}f_{ajk}f_{bjk}=N\, \delta_{ab}~.
}

In the enveloping algebra implicit in the fundamental representation, we have
also
\eq{
\label{anticom:F}
\{F_i,F_j\}=\mn{\frac{1}{2N}}\delta_{ij}\id+d_{ijk}F_k~.
}
The coefficients $d_{ijk}$ are totally symmetric under exchange of indices and
they are familiar for SU(3)\;\cite{IZ}.
These coefficients can be obtained from the fundamental representation
\eq{
\label{dijk}
d_{ijk}=2\Tr[\{F_i,F_j\}F_k]
=\mn{\frac{1}{4}}\Tr[\{\lambda_i,\lambda_j\}\lambda_k]
~,
}
and obey the property
\eq{
\label{norm:dijk}
\sum_{j,k=1}^{d}d_{ajk}d_{bjk}=\frac{N^2-4}{N}\, \delta_{ab}~.
}

Taking the trace of Eq.\;\eqref{anticom:F} we obtain the value of the second
order Casimir invariant
\eq{
\label{casimir:F}
\sum_{i=1}^{d}(F_i)^2=C_2(F)\id~,
}
where $C_2(F)=\frac{d}{2N}$. The second order Casimir invariant for the adjoint
representation is already given by Eq.\;\eqref{norm:fijk} which imply
$C_2(\ad)=N$.

\subsection{The fundamental representation for $SU(N)$}
\label{app:sun:F}

We show here a explicit choice of matrices for the fundamental
representation of $SU(N)$ in the Cartan-Weyl basis. With certain choice of
phases and cocycles implicit, such choice coincides with the Gell-Mann type
matrices (except for a factor one-half).

The $SU(N)$ algebra $\mathcal{G}$ is the algebra of the hermitian and traceless
$N\times N$ matrices. This is the defining and a fundamental (and minimal)
representation. An orthogonal basis for this algebra can be chosen to be the
$d$ matrices
\eqarr{
\label{hi}
h_k&=&\mn{\frac{\ms{1}}{\ms{\sqrt{2k(k+1)}}}}\,\diag(\id_k,-k,0,\ldots,0)~,
~~k=1,\ldots, r, \\
\label{sa}
S_{ij}&=&\mn{\frac{1}{2}}(e_{ij}+e_{ji})~,~~i<j=1,\ldots,N,
\\
\label{aa}
A_{ij}&=&\mn{\frac{1}{2i}}(e_{ij}-e_{ji})~,~~i<j=1,\ldots,N,
}
where $r=N-1$, and $e_{ij}$ denotes the canonical basis defined by
$(e_{ij})_{kl} = \delta_{ik}\delta_{jl}$.
Each type of matrices spans the algebra subspaces $\{h_k\}\sim\ft_r$,
$\{S_{ij}\}\sim\tilde{\ft}_q$ and $\{A_{ij}\}\sim\ft_q$ in
Eq.\;\eqref{ordering} and the normalization satisfies Eq.\;\eqref{norm:F}. If we
associate $(i,j)=(i,i+1)\leftrightarrow \alpha_i$ [$i=1,\ldots,r$],
$(i,j)=(i,i+2)\leftrightarrow \alpha_{r+i}$ [$i=1,\ldots,r-1$], $\ldots$,
$(1,N)\leftrightarrow \alpha_q$, we obtain the correspondence
$S_{ij}\leftrightarrow \cs_\alpha$ and $A_{ij}\leftrightarrow \ca_\alpha$;
$q=(d-r)/2=N(N-1)/2$ is the number of positive roots of the algebra denoted by
$\alpha$, used for labeling $\cs_\alpha$ and $\ca_\alpha$. The first $r$ roots
are the simple roots. All positive roots can be written as combinations of the
simple roots. Since $SU(N)$ is a simply laced algebra\;\cite{slansky}, the
positive roots are given, in terms of the simple roots,
\eqarr{
\text{h=1}&&\alpha_1,\alpha_2,\ldots,\alpha_r~,\cr
\text{h=2}&&\alpha_1+\alpha_2,\alpha_2+\alpha_3,\ldots,\alpha_{r-1}
+\alpha_r ~,\cr
\text{h=3}&&\alpha_1+\alpha_2+\alpha_3,\alpha_2+\alpha_3+\alpha_4,\ldots,
\alpha_{r-2}+\alpha_{r-1}+\alpha_r ~,\cr
\vdots\hs{1em}&&\hs{1em}\vdots\cr
\text{h=r}&&\alpha_1+\alpha_2+\ldots+\alpha_r~.
}
The height $h$ is the sum of the expansion coefficients of the positive roots in
terms of the simple roots. Only the sum of neighbor simple roots are also roots.

The roots live in an Euclidean $r$-dimensional space and explicit coordinates can
be obtained from the matrices \eqref{hi}, \eqref{sa} and \eqref{aa}, and the
relation
\eq{
\label{hE}
[h_k,E_{\alpha}]=(\alpha)_k\,E_{\alpha}~.
}
Using $E_{\alpha}=e_{i,i+1}$, the simple roots $\alpha_i$, which are normalized
as $(\alpha_i,\alpha_i)=1$, have coordinates
\eq{
(\alpha_i)_k=(h_k)_{ii}-(h_k)_{i+1,i+1}~.
}

The weight system of the fundamental representation have highest weight
$\lambda_1$, which is just the first primitive weight defined by
$2(\lambda_i,\alpha_j)/(\alpha_j,\alpha_j)=\delta_{ij}$, $i,j=1,\ldots,r$.
The $r+1$ weights of this representation can be obtained by subtracting
positive roots from the highest weight:
\eq{
\label{pesos:F}
\begin{array}{rll}
\mu_0=&\lambda_1&\sim (10\ldots 0)\cr
\mu_1=&\lambda_1-\alpha_1&\sim (\ms{-}110\ldots 0)\cr
\mu_2=&\lambda_1-\alpha_1-\alpha_2&\sim (0\ms{-}110\ldots 0)\cr
\vdots \hs{1.4em}&\vdots& \hs{3em}\vdots\cr
\mu_r=&\lambda_1-\alpha_1-\cdots-\alpha_r&\sim (0\ldots 0\ms{-}1)~.
\end{array}
}
The last column corresponds to the weights in Dynkin basis\;\cite{slansky}
$\mu_a\sim (n_1n_2\ldots n_r)$, which are the expansion coefficients in terms of
the primitive weights $\mu_a=\sum_{i=1}^{r}n_i\lambda_i$. These $r+1=N$ weights
can label all the states, which are not degenerate in this case.

The matrices \eqref{hi} represents the Cartan subalgebra, and in the
Cartan-Weyl basis  they are diagonal. The diagonal elements are just the
components of the weights in Eq.\;\eqref{pesos:F}, i.e.,
\eq{
h_k\equiv \bra{\mu}h_k\ket{\mu'}=
\diag(\mu_0,\mu_1,\ldots,\mu_r)_k~.
}

\subsection{$f_{ijk}$ and $d_{ijk}$ tensors}
\label{f,d:ijk}

By using Eqs.\;\eqref{fijk}, \eqref{dijk} and the properties of the fundamental
representation described in the preceding subsection, we can deduce some general
features of the rank-3 tensors $f_{ijk}$ and $d_{ijk}$ with the ordering defined
by Eq.\;\eqref{ordering}. Firstly, we define \eqarr{
f(F_i,F_j,F_k)&\equiv& f_{ijk}~,\\
d(F_i,F_j,F_k)&\equiv& d_{ijk}~.
}
Then, the following properties can be proved,
\eqarr{
\label{fhi}
f(h_i,h_j,F_k)&=&0\cr
f(h_i,\cs_\alpha,\cs_\beta)&=&0\cr
f(h_i,\ca_\alpha,\ca_\beta)&=&0\cr
f(h_i,\ca_\alpha,\cs_\beta)&=&-(\alpha)_i\,\delta_{\alpha\beta}\cr
f(h_i,\cs_\alpha,\ca_\beta)&=&(\alpha)_i\,\delta_{\alpha\beta}~.
}
The zeros above can be obtained from the general relations
\eqarr{
\label{tpq:a}
[\ft_p,\ft_p]&\subset &\ft_q\cr
[\ft_q,\ft_q]&\subset &\ft_q\cr
[\ft_q,\ft_p]&\subset &\ft_p~,
}
and
\eqarr{
[\ft_r,\ft_r]&= &0\cr
[\ft_r,\ft_q]&\subset &\tilde{\ft}_q\cr
[\ft_r,\tilde{\ft}_q]&\subset &\ft_q~,
}
and the fact that the trace of the product of two elements of distinct
subspaces is null (orthogonality). The properties \eqref{tpq:a} are easily seen
in the fundamental representation by the symmetric character of the elements of
$\ft_p$, the antisymmetric character of the elements of $\ft_q$, and the
properties of a commutator $[A,B]^{\tp}=-[A^{\tp},B^{\tp}]$.

For illustration, we will show how to obtain the non-null elements of
Eq.\;\eqref{fhi}, knowing the properties of the fundamental representation in
the Cartan-Weyl basis. The procedure is as follows,
\eqarr{
\label{exemplo1}
f(h_i,\ca_\alpha,\cs_\beta)&=&\mn{\frac{2}{i}}\sum_{\mu}
\bra{\mu}[h_i,\ca_\alpha]\cs_\beta\ket{\mu}\cr
&=&-2(\alpha)_i\sum_{\mu}\bra{\mu}\cs_\alpha\cs_\beta\ket{\mu}\cr
&=&-\mn{\frac{1}{2}}(\alpha)_i\sum_{\mu}\bra{\mu}E_{\alpha}E_{-\beta}
+E_{-\alpha}E_\beta\ket{\mu}\cr
&=&-(\alpha)_i\delta_{\alpha\beta}\mn{\frac{1}{2}}
\sum_{\mu}[\delta(\mu-\alpha)+\delta(\mu+\alpha)]~,
}
where
\eq{
\delta(\mu)=\left\{
\begin{array}{l}
1~, ~~\text{$\mu$ is a weight,}\cr
0~, ~~\text{$\mu$ is not a weight.}
\end{array}\right.
}
In addition, we have used Eqs.\;\eqref{cs}, \eqref{ca}, \eqref{hE} and
$\bra{\mu}E_{-\alpha}E_\alpha\ket{\mu}=\delta(\mu+\alpha)$. We can see from
Eq.\;\eqref{pesos:F} that the last sum of Eq.\;\eqref{exemplo1} gives $2$ for
any positive root $\alpha$ since there are always one positive root
connecting two weights.

For $d_{ijk}$, the fundamental representation is essential since we can not use
the Lie algebra properties. Some properties are
\eqarr{
d(h_i,h_j,\cs_\alpha)&=&d(h_i,h_j,\ca_\alpha)=0\cr
d(h_i,\cs_\alpha,\ca_\beta)&=&0\cr
d(h_i,\cs_\alpha,\cs_\beta)&=&\delta_{\alpha\beta}
\sum_{\mu}(\mu)_i[\delta(\mu-\alpha)+\delta(\mu+\alpha)]\cr
d(h_i,\ca_\alpha,\ca_\beta)&=&d(h_i,\cs_\alpha,\cs_\beta)\cr
d(h_i,h_j,h_k)&=&4\sum_{\mu}(\mu)_i(\mu)_j(\mu)_k~.
}
Basis independent properties can be extracted by defining a symmetric
algebra\;\cite{ORaif,michel} in the space of  the $N\times N$
tracesless hermitian matrices. Such space will be denoted by
$\mathcal{M}_h(N,\mathbb{C})$, and it is isomorphic to a $\mathbb{R}^d$ vector
space. Given $\undertilde{x},\undertilde{y}\in\mathcal{M}_h(N,\mathbb{C})$, the
symmetric algebra is defined as
\eq{
\label{alg:s}
\undertilde{x}\vee\undertilde{y}\equiv
\mn{\frac{1}{2}}\{\undertilde{x},\undertilde{y}\}
-\mn{\frac{1}{N}}\Tr[\undertilde{x}\undertilde{y}]~. }
Obviously $\undertilde{x}\vee\undertilde{y} \in \mathcal{M}_h(N,\mathbb{C})$.
The tilde in $\undertilde{x}$ means
\eq{
\label{x->tx}
\undertilde{x}\equiv \bx\cdot\bs{\lambda}=x_i\lambda_i~,
}
where $\bx$ lives in $\mathbb{R}^d$, in the adjoint representation space.
In terms of the vectors $\bx$ and $\mathbf{y}$ in the adjoint, the symmetric
algebra \eqref{alg:s} can be written
\eq{
\label{alg->prod:s}
\undertilde{x}\vee\undertilde{y}\equiv (\bx\vee\mathbf{y})\cdot\bs{\lambda}~,
}
where the $\vee$ in the righthand side of Eq.\;\eqref{alg->prod:s} is the
product defined on the adjoint vectors, Eq.\;\eqref{prod:s}. We use the same
symbol for both of the products.

From the symmetric and antisymmetric nature of the elements of $\ft_p$ and
$\ft_q$, respectively, we can conclude that
\eqarr{
\label{tpq:s}
\ft_p\vee\ft_p &\subset& \ft_p \cr
\ft_q\vee\ft_q &\subset& \ft_p \cr
\ft_q\vee\ft_p &\subset& \ft_q~.
}
Then, the equality $d_{ijk}=2\Tr[\{F_i,F_j\}F_k]=4\Tr[F_i\vee F_jF_k]$ yields
\eqarr{
d(\ft_p,\ft_p,\ft_q)&=&0\cr
d(\ft_q,\ft_q,\ft_q)&=&0~.
}
The results above are invariant if the positions of any pair are interchanged.
For particular elements of $\cG$, we also have $d(F_i,F_i,F_i)=0$, for
$F_i=\cs_\alpha$ and $F_i=h_1$. However, any element $F$ in $\tilde{\ft}_q$ does
not satisfy $d(F,F,F)=0$, differently of $\ft_q$. Thus $\ft_q$ forms a 
subspace (and subalgebra) of $\cG$ but not $\tilde{\ft}_q$.

\section{Branching of adj$SU(N)$ with respect to real $SO(N)$}
\label{app:d=p+q}

The separation of the $SU(N)$ algebra in $\ft_p$ and $\ft_q$, as in
Eq.\;\eqref{ordering}, naturally induces two representations of the $SO(N)$
subalgebra generated by $\ft_q=\{\ca_\alpha\}$.

One of them is just the adjoint of $SO(N)$ carried by the real
antisymmetric $N\times N$ matrices spanned by $\{i\ca_\alpha\}$, for which the
subgroup action is
\eq{
e^{i\theta_\alpha \ca_\alpha} (a_\beta i\ca_\beta)
e^{-i\theta_\alpha\ca_\alpha} =
i\ca_\alpha D_1(e^{i\bs{\theta}\cdot\bs{\ca}})_{\alpha\beta} a_\beta
~.
}
The representation $D_1$ is just the lower $q\times q$ block of the adjoint
representation of $\exp(\theta_\alpha D(i\ca_\alpha))$ of $SU(N)$ with the
ordering \eqref{ordering}. This is an irrep of dimension $q=N(N-1)/2$.

The other representation is carried by the real $N\times N$ symmetric traceless
matrices spanned by $\ft_p=\{h_i,S_\alpha\}$. The subgroup action is given by
\eq{
e^{i\theta_\alpha \ca_\alpha} (a_i h_i+b_\beta S_\beta)
e^{-i\theta_\alpha\ca_\alpha} =
(h_i, S_\alpha)
D_2(e^{i\bs{\theta}\cdot\bs{\ca}})
\begin{pmatrix}
a_i \cr b_\beta
\end{pmatrix}
~.
}
The representation $D_2$ is just the upper $p\times p$ block matrix in the
adjoint representation $\exp(\theta_\alpha D(iA_\alpha))$ of $SU(N)$ whose
dimension is $p=N(N+1)/2-1$ and it is irreducible.

\section{Properties of matrices similar to antisymmetric matrices}
\label{app:A}

The following proposition will be proved: For any complex or real $n\times n$
diagonalizable\,\cite{endnote4} matrix $X$,
\eq{
\label{tr^k=0}
\Tr[X^{2m+1}]=0~~~\text{for all $2m+1\le n$}
}
imply $X$ is similar to an antisymmetric matrix $A=UXU^{-1}$. The converse is
trivial since the trace of a matrix is equal to the trace of the transpose.

For the proof, we need the characteristic equation\;\cite{michel}
\eq{
\det(X-\lambda \id)=(-1)^n
[\lambda^n-\sum_{k=1}^{n}\gamma_k(X)\lambda^{n-k}]
~,
}
where
\eqarr{
\gamma_1(X)&=&\Tr[X]~,\\
\gamma_2(X)&=&\ml{\frac{1}{2}}\Tr[X^2-\gamma_1(X)X]~,\\
\gamma_3(X)&=&\ml{\frac{1}{3}}\Tr[X^3-\gamma_1(X)X^2-\gamma_2(X)X]~,\\
&\vdots&\\
\gamma_n(X)&=&\ml{\frac{1}{n}}\Tr[X^n-\sum_{k=1}^{n-1}\gamma_k(X)X^{n-k}]
~.
}
The same coefficients enter in the matricial equation
\eq{
\label{x^k}
X^n-\sum_{k=1}^{n}\gamma_k(X)X^{n-k}=0~,
}
for which $X^0=\id_n$ is implicit.

If Eq.\;\eqref{tr^k=0} is satisfied, all odd coefficients $\gamma_{2k+1}(X)=0$
and the characteristic equation reads
\eq{
\label{detx2}
\det(X-\lambda\id)=(-1)^n
[\lambda^n-\gamma_2(X)\lambda^{n-2}-\gamma_4(X)\lambda^{n-4}-\ldots-\gamma_n(X)]
~.
}
If $n$ even we rewrite $n=2m$ and Eq.\;\eqref{detx2} yields
\eq{
\label{detx2:even}
\det(X-\lambda\id)=(\lambda^2)^m-\sum_{k=1}^m\gamma_{2k}(X)(\lambda^2)^{m-k}
=f(\lambda^2) ~.
}
If $n$ odd we rewrite $n=2m+1$ and Eq.\;\eqref{detx2} yields
\eq{
\label{detx2:odd}
\det(X-\lambda\id)=-\lambda
[(\lambda^2)^m-\sum_{k=1}^m\gamma_{2k}(X)(\lambda^2)^{m-k}]
=-\lambda f(\lambda^2)~.
}
For both Eqs.\;\eqref{detx2:even} and \eqref{detx2:odd}, $f(\lambda^2)$ is
a polynomial in $\lambda^2$ of order $m$ and it has, including degeneracies, $m$
(complex) roots $\lambda_i^2$, $i=1,2,\ldots,m$. Then
\eq{
f(\lambda^2)=\prod_{i=1}^{m}(\lambda^2-\lambda^2_i)=
\prod_{i=1}^{m}(\lambda-\lambda_i)(\lambda+\lambda_i)
~,
}
which implies that for each eigenvalue $\lambda_i$ of $X$  an opposite
eigenvalue $-\lambda_i$ exists (both might be zero), except for a unique
additional zero eigenvalue when $n$ is odd, as can be seen from
Eq.\;\eqref{detx2:odd}.

The existence of opposite eigenvalues guarantees the existence of a similarity
transformation $U_1$ that leads $X$ to the diagonal form
\eq{
\label{U1}
U_1XU_1^{-1}=\left\{
\begin{array}{ll}
\diag(\lambda_1\sigma_3,\lambda_2\sigma_3,\ldots,\lambda_m\sigma_3)
& \quad\text{for $n=2m$,}
\cr
\diag(\lambda_1\sigma_3,\lambda_2\sigma_3,\ldots,\lambda_m\sigma_3,0)
& \quad\text{for $n=2m+1$.}
\end{array}\right.
}
Then, one can use the matrix
\eq{
U_2=\left\{
\begin{array}{ll}
\id_m\otimes e^{-i\sigma_1\pi/4}
& \quad\text{for $n=2m$,}
\cr
\diag(\id_m\otimes e^{-i\sigma_1\pi/4},0)
& \quad\text{for $n=2m+1$,}
\end{array}\right.
}
to transform Eq.\;\eqref{U1} into antisymmetric form
\eq{
U_2U_1XU_1^{-1}U_2^{-1}=
\left\{
\begin{array}{ll}
\diag(\lambda_1\sigma_2,\lambda_2\sigma_2,\ldots,\lambda_m\sigma_2)
& \quad\text{for $n=2m$,}
\cr
\diag(\lambda_1\sigma_2,\lambda_2\sigma_2,\ldots,\lambda_m\sigma_2,0)
& \quad\text{for $n=2m+1$.}
\end{array}\right.
}
When $X$ is hermitian, $U_1$ can be unitary and the eigenvalues $\lambda_i$ are
real. For $X$ in the $SU(N)$ algebra, condition \eqref{tr^k=0} is necessary and
sufficient for $X$ to be in the orbit of an element in $\ft_q$.

\section{The decomposition of $(d\otimes d)_S$ of $SU(N)$}
\label{dxd}

 In the last term of Eq.\;\eqref{V3:N>2}, $\tilde{\Lambda}$ transforms under
the $(d\otimes d)_S$ representation of $SU(N)_H$. Such representation is
reducible. Though, unlike the $N=2$ case, it has more than two
components, as shown in the table\;\ref{tab:sun} for $N=2,\ldots,6$. (It can be
proved that the number of components are at most four.) Table\;\ref{tab:sun}
shows the branchings of the direct product representation $d\otimes d$ for the
symmetric part denoted by the subscript $S$\;\cite{slansky}; the last column
shows the dimension of the representation space of $(d\otimes d)_S$, which is
just the space of the real symmetric $d\times d$ matrices.

\begin{table}[htb]
\caption[]{$SU(N)$ decompositions
\label{tab:sun}}
\begin{center}
\begin{tabular}{|c||c|c|c|}
\hline
$~N~$ & $d=N^2-1$ 
& $ (d\otimes d)_{\rm S}$ &$\frac{d(d+1)}{2}$  \\
\hline
2 &\bf 3  
& $\bf 5\oplus 1$\bf  & 6  \\
\hline
3 &\bf 8 
& $\bf  27\oplus 8\oplus 1$\bf  & 36  \\
\hline
4 &\bf 15 
& $\bf 84\oplus 20\oplus 15\oplus 1$\bf  & 120  \\
\hline
5 &\bf 24 
& $\bf  200\oplus 75 \oplus 24\oplus 1$\bf  & 300  \\
\hline
6 &\bf 35 
& $\bf 405\oplus 189 \oplus 35\oplus 1$\bf  & 630  \\
\hline
\end{tabular}
\end{center}
\end{table}

\section{Real symmetric adjoint representation in $(d\otimes d)_S$}
\label{app:adj:S}

We know the adjoint representation for a Lie group can be obtained from the
vector space spanned by the algebra itself in any representation. In particular,
any compact semisimple Lie algebra can be represented by the  $d\times d$
real antisymmetric matrices given by the structure constants
$i(T_i)_{jk}=f_{ijk}$.

In contrast, there is also a $d\times d$ real symmetric representation spanned
by the real symmetric matrices $\{d_i\}$ given by
\eq{
\label{adjS:rep}
(d_i)_{jk}=d_{ijk}~,
}
which is the rank-3 totally symmetric tensor from Eq.\,\eqref{dijk}.

That $\{d_i\}$ represents the $SU(N)$ in the adjoint representation can be seen
by
\eq{
\label{adjS:alg}
[T_i,d_j]=if_{ijk}d_k~.
}
Equation \eqref{adjS:alg} can be proved by using Eqs.\;\eqref{fijk},
\eqref{dijk} and the completeness relation of Eq.\;\eqref{complete}.
Moreover
\eq{
d_{ijk}T_jT_k= -\frac{N}{2}d_i~.
}

Thus, for $N\ge 3$, the component in the adjoint of the tensor $\tilde{\Lambda}$
can be extracted as
\eq{
\label{L:ad}
\tilde{\Lambda}\big{|}_{\mt{\ad}}=\tilde{\Lambda}^{\mt{(\ad)}}_i d_i~,
}
where, from Eq.\;\eqref{norm:dijk},
\eq{
\tilde{\Lambda}^{\mt{(\ad)}}_i=\mn{\frac{N}{N^2-4}}\Tr[\tilde{\Lambda}d_i]
=\mn{\frac{N}{N^2-4}}d_{ijk}\tilde{\Lambda}_{jk}
~.
}
This is a practical way of extracting the symmetric adjoint representation of
$(d\otimes d)_S$.

\acknowledgments
This work was supported by Conselho Nacional de Desenvolvimento
Cient\'\i fico e Tecnol\'ogico (CNPq). The author would like to thank Prof.
J. C. Montero and Prof. V. Pleitez for pointing up Refs.\;\onlinecite{Haber.2}
and \onlinecite{GH} which motivated this work, and Prof. L. A. Ferreira
for general discussions on Lie algebras and Lie groups.



\begin{thebibliography}{99}
\bibitem{endnote0}
Note that this definition differs from Ref.\,\onlinecite{GH}.
\bibitem{endnote1}
There can be the quartic vector-vector interaction
$(\Phi_a^\dag\vec{\tau}\Phi_a).(\Phi_b^\dag\vec{\tau}\Phi_b)$
(see, {\it e.g.}, Ref.\;\onlinecite{Weinberg}) but it can be reduced to the form
\eqref{V:N=2} by using the completeness relation Eq.\;\eqref{complete} for the
generators of $SU(2)_L$.
\bibitem{endnote3}
Generalized, in the sense that the number of reflected directions in the
CP-reflection \eqref{reflection:N>2} is $q$ which can be even.
The adjoint representation space is isomorphic to $\mathbb{R}^d$
and true pseudoscalars should be invariant by $SO(d)$ but should change sign
under any reflection of an odd number of directions.
\bibitem{endnote4}
This means that in the canonical Jordan form $X$ does not have any
off-diagonal element different from zero.
\end{thebibliography}
\end{document}